\documentclass{jfm}
\usepackage{graphicx}
\usepackage{psfrag}
\usepackage{natbib}
\usepackage{amsmath}
\usepackage{amssymb}
\usepackage{color}
\usepackage{soul}
\usepackage{subfigure, wrapfig, psfrag} 
\usepackage{mathtools,dsfont}
\usepackage{bm} 
\usepackage{pifont}

\newcommand{\beq}{\begin{equation} }
\newcommand{\eeq}{\end{equation} }

\newcommand\Real{\mbox{Re}} 
\newcommand\Imag{\mbox{Im}} 

\newcommand*{\D} {\text{d}}
\newcommand*{\dt} {\text{dt}}
\newcommand*{\dtau} {\text{d}\tau}

\newcommand*{\lec} {\ell_{ec}}
\newcommand*{\kec} {K}

\newcommand*{\zh} {\hat{\eta}}
\newcommand*{\jj} {\mathrm{i}}
\newcommand*{\ith} {j^\mathrm{th}}

\newcommand*{\ccr} {\tilde{c}_\mathrm{cr}}
\newcommand*{\ct} {\tilde{c}}

\newcommand*{\Np} {N_{p}}
\newcommand*{\Npmax} {N_p^{\mathrm{max}}}
\newcommand*{\Nmean} {\langle \mathcal{N} \rangle}
\newcommand*{\Nmax} {\mathcal{N}_\mathrm{max}}
\newcommand*{\Ntot} {N}

\newcommand*{\Ns} {\mathcal{N}}
\newcommand*{\xo} {x_0}
\newcommand*{\Pc} {\mathcal{P}}
\newcommand*{\ths} {\theta^*}
\newcommand{\upd} {\mathrm{d}}
\newcommand{\gjk}{g_{j,k}}  
\newcommand{\cmax} {c_{max}}

\usepackage{ifpdf}
\usepackage{graphics}

\ifpdf
    \DeclareGraphicsExtensions{.png, .jpg, .pdf}
\graphicspath{
{./Figs/}
{./ModifiedFigs/}
{./ProductionFiles/}
}
\fi
 
\title{A fluid-mechanical model of elastocapillary coalescence}

\author[K. Singh, J. R. Lister \& D. Vella]%
{K\ls I\ls R\ls A\ls N\ns S\ls I\ls N\ls G\ls H$^1$,%
\ns
 J\ls O\ls H\ls N\ns R. \ns L\ls I\ls S\ls T\ls E\ls R$^2$
\and D\ls O\ls M\ls I\ls N\ls I\ls C\ns V\ls E\ls L\ls L\ls A$^1$
\thanks{Email address for correspondence: dominic.vella@maths.ox.ac.uk}
}
\affiliation{$^{1}${OCCAM, Mathematical Institute, Radcliffe Observatory Quarter,\\Woodstock Road, Oxford, OX2 6GG, UK}\\
$^{2}${Department of Applied Mathematics and Theoretical Physics,\\ Centre for Mathematical Sciences, Wilberforce Road, Cambridge, CB3 0WA, UK}
}
\date{\today}

\begin{document} 

\maketitle

\begin{abstract}
We present a fluid-mechanical model of the coalescence of a number of elastic objects due to surface tension. We consider an array of spring-block elements separated by thin liquid films, whose dynamics are modelled using lubrication theory. With this simplified model of elastocapillary coalescence, we present the results of numerical simulations for a large number of elements, $N=O(10^4)$. A linear stability analysis shows that pairwise coalescence is always the most unstable mode of deformation. However, the numerical simulations show that the cluster sizes actually produced by coalescence from a small white-noise perturbation have a distribution that depends on the relative strength of surface tension and elasticity, as measured by an elastocapillary number $K$. Both the maximum cluster size and the mean cluster size scale like $K^{-1/2}$ for small $K$. An analytical solution for the response of the system to a localized perturbation shows that such perturbations generate propagating disturbance fronts, which leave behind `frozen-in' clusters of a predictable size that also depends on $K$. A good quantitative comparison between the cluster-size statistics from noisy perturbations and this `frozen-in' cluster size suggests that  propagating fronts may play a crucial role in the dynamics of coalescence.
\end{abstract}

\section{Introduction}
\label{sec:introduction}

\begin{figure}
\centering
\includegraphics[width=0.975\textwidth,angle=0]{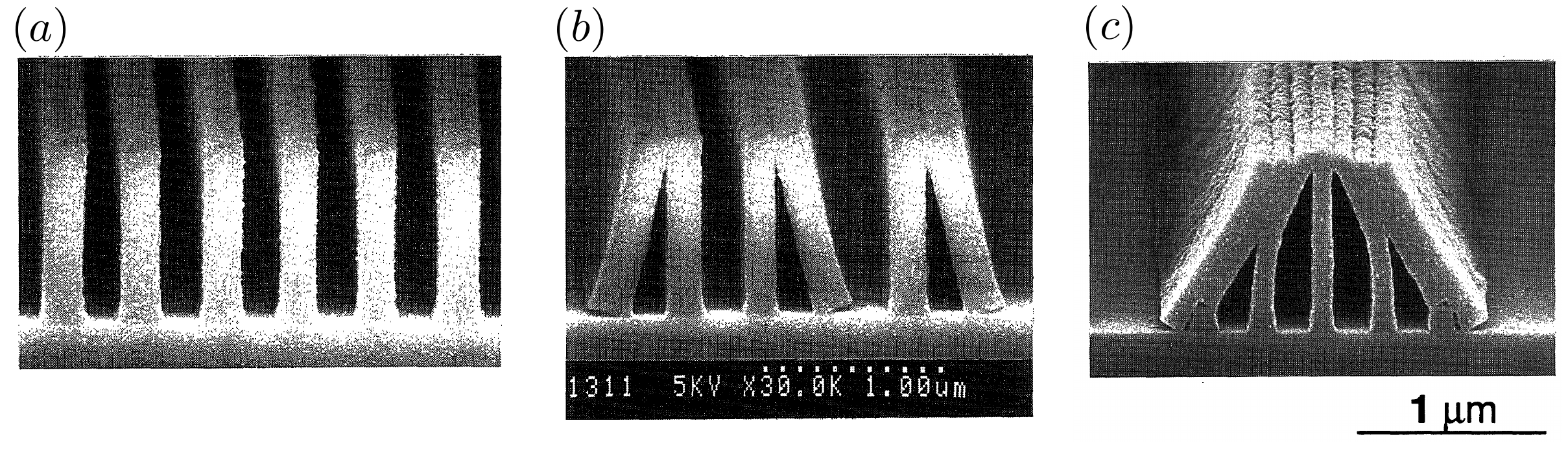}
\caption{Scanning electron micrographs show microscopic patterns formed by a technique common to the microlithography used in the manufacture of MEMS devices: micropillar arrays are formed by UV exposure of a photoresist, which is subsequently rinsed and the solvent evaporated. During this second phase, surface tension forces can cause individual elements to come into contact and subsequently stick (via van der Waals forces). In (a) a low surface tension liquid (a $50:50$ mixture of tert-butylalcohol and water) is used in the rinse phase and no deformation is observed. In (b) $100\%$ water is used and the resist collapses into pairs. (The scale bar is the same in (a) and (b).) (c) A small aspect ratio pattern collapses into bundles with a larger characteristic size. (Images reproduced from \cite{Tanaka1993}, $\copyright$ 1993 {\it  Japan Society of Applied Physics} with permission).
}
\label{fig:pics}
\end{figure}

Elastocapillary effects occur, by definition, when capillary forces are of the same order as the elastic stresses required to deform a structure. 
While elastocapillarity may be observed in everyday life and at macroscopic length scales, e.g.~the clumping of paintbrush fibres wetted by a liquid \citep{Bico2004}, macroscopic structures are usually built to withstand gravity and are insufficiently compliant to be significantly deformed by surface tension.  At microscopic scales, surface-tension forces become extremely important, and so elastocapillary effects become ubiquitous, and cause significant deformations as they pull structures together: as examples, highly-ordered bundle structures can form from an array of nano-pillars in contact with an evaporating liquid \cite[][]{Chakrapani2004,Pokroy2009}, carbon nanotubes can be `densified' \cite[][]{devolder2011,devolder2013}, and the details of how oil droplets wet and deform the microscopic barbs and barbules of marine bird feathers is one of the reasons that oil slicks are so damaging to wildlife \citep{OHara2010,Duprat2012}.  Indeed, the fact that surface tension can bend, and even break, micro-structures is an important consideration in the wet etching process of MEMS fabrication \citep{Tanaka1993}. Figure~\ref{fig:pics} illustrates an example of  elastocapillary-induced sticking of structures formed in microlithography.

The surface tension of an interface, $\gamma$, acts to minimize the interfacial area given any mechanical constraints; the bending stiffness of an object, denoted $B$, acts to minimize the deflection of an elastic object from its undeformed state. When these two forces act in opposition, the deformation that results depends on the typical size of the system compared to an \emph{elastocapillary length} $\lec=(B/\gamma)^{1/2}$ \cite[][]{Bico2004}. Heuristically speaking, when the typical system size is larger than $\lec$ then the elastic object will be deformed by surface tension; when the typical system size is smaller than $\lec$ then the object will remain undeformed \cite[][]{Neukirch2007}. 

While the deflection of an individual elastic object by surface tension is of some academic interest \cite[][]{Neukirch2007}, and is relevant to the locomotion of insects and robots on water \cite[][]{Song2007,Vella2008,Bush2008}, the elastocapillary interaction between multiple elastic objects is  both a rich source of fluid-mechanical problems and relevant for more applications. For example, during capillary rise in a single channel with flexible walls, the surface tension--induced bending of the channel walls causes the effective channel thickness to decrease and results in an equilibrium rise height that is larger than the classic Jurin height \cite[][]{Kim2006}. With capillary rise into the channels between an array of many plates or fibres, a complex hierarchical structure results with clusters of various sizes emerging and indeed clusters that fragment along their length \cite[][]{Bico2004}. This intriguing phenomenon is governed by an interplay between three length scales: the length of each element, the spacing between elements and the elastocapillary length, $\lec$. The reverse problem, that of a one-dimensional array of fibres that is initially submerged beneath an interface but buckles as the fibres pierce the liquid interface, was considered by \cite{Chiodi2010}. Although some experimental studies have suggested that evaporation of the liquid is responsible for coalescence \citep{Tanaka1993,deLangre2009,Pokroy2009}, other elastocapillary systems show that coalescence can occur without evaporation \cite[][]{Bico2004,Py2007b,Boud2007,Gat2013}. In each of these various scenarios, the mechanics governing the distribution of final cluster size have been discussed but are far from trivial \cite[][]{Py2007b,Boud2007}. 

The phenomena already described have largely been studied in one-dimensional scenarios. However, two-dimensional arrays of fibres exhibit  a richer phenomenology including the breaking of chiral symmetry \cite[][]{Pokroy2009}, while two-dimensional effects can cause capillary rise in an elastic Hele--Shaw cell to continue indefinitely \cite*[][]{Cambau2011}. Several studies have also demonstrated the deformation of an elastic membrane by liquid droplets. For example, thin elastic sheets can be wrapped up spontaneously by the presence of a liquid droplet \cite[][]{Py2007} while extremely thin membranes wrinkle in similar scenarios \citep{Huang2007,Vella2010,Schroll2013}. 

The majority of previous work focusses on understanding the equilibrium of elastic objects subject to capillary forces. More recently, attention has shifted to understanding the timescales and dynamic properties of liquid motion in elastocapillarity. Given the small length scales that are usually involved in elastocapillary phenomena, the typically Reynolds numbers involved are usually small. Motivated by the long, thin geometry in elastocapillary rise, \cite{Aristoff2011} and \cite{Duprat2011} developed a fluid model based on lubrication theory for the elastocapillary dynamics between two flexible elements open to an infinite bath of liquid. Their approach led to a sixth-order nonlinear diffusion equation for the gap thickness, which they solved numerically. \cite{Taroni2012} developed a similar model to describe the approach to equilibrium of a pair of elastic beams deformed by a liquid drop with a given volume. In this system up to four equilibria may exist for given material parameters; the dynamic calculation allowed the stability of these equilibria to be determined as well as the growth rates of instability.

Recently, \cite{Gat2013} studied the dynamics of elastocapillary coalescence in an array of elastic sheets that trap a finite volume of liquid. They used a lubrication-type model to describe the fluid flow and aggregation of elements, albeit using a linearized relationship between the displacement of the elastic elements and the fluid force applied on them. A linear stability analysis of their model equations showed that pairwise clustering is the fastest growing mode of instability. Iterating this pairwise collapse, they deduced that the system  moves through a series of states in which clusters of size $2^n$ are formed, ending only when the clusters are so large that they are stable to further coalescence.

In this article, we develop a model similar in spirit to that presented by \cite{Gat2013}. We consider a series of elastic elements, characterized by  simple displacements proportional to the net force exerted on them, and subject to the fluid forces that arise from a lubrication-type flow and surface tension. The simplicity of our elastic model allows us to consider systems containing large numbers of elements ($\Ntot\sim 10^4$) computationally, and thereby to perform a detailed study of clustering dynamics in large systems. The competition between the elasticity of the structure and surface tension  is characterized by an `elastocapillary number' \citep[][]{Mastrangelo1993}, which is in effect a dimensionless spring stiffness. We go beyond the linear stability analysis of \cite{Gat2013} by presenting numerical solutions of the nonlinear governing equations that result, as well as studying the response of the system to a localized perturbation, for various values of the dimensionless spring stiffness. For relatively large spring stiffnesses, the system is stable and no clustering is observed. However, for sufficiently small spring stiffnesses the system is unstable to clustering, and the dynamics are considerably more complex than pairwise collapse; we study the cluster statistics that emerge from a random initial condition as the spring stiffness varies.

A novel feature of our results is that they suggest that the effects of small local perturbations have been overlooked by previous studies. For sufficiently small spring stiffnesses, the system is an intrinsically unstable medium in which localized perturbations lead to the propagation of fronts through the system. This makes the phenomenology of dynamic coalescence similar to that observed in other systems involving front propagation into an unstable medium such as the pearling instability of tubular vesicles \cite[][]{Goldstein1996,Powers1997} and the initiation and propagation of drops in the thin-film Rayleigh--Taylor instability \cite[][]{Fermigier,Limat1992} to mention two such systems \cite[see][for a review of this class of problems]{vanS2003}. We calculate the speed of propagation of the coalescence front as well as the typical cluster size that forms behind the front and find that this cluster size  generally gives an excellent estimate of the maximum cluster size that is observed in our detailed simulations. This suggests that front propagation due to localized imperfections may be the physical mechanism behind previous observations that random imperfections induce a broad distribution of cluster sizes \cite[][]{Boud2007}. However, our results also suggest that the distribution of cluster sizes, once rescaled by the mean cluster size, follows a universal Gaussian distribution. 

The plan of this article is as follows. In \S\,\ref{sec:model} we present our model of multi-body elastocapillary dynamics and then consider an illustrative reduction to two interacting elastic elements and present some illustrative numerical simulations for large numbers of elements. In \S\,\ref{sec:linstab} we analyse the linear stability of the system both for an array of discrete elements and for the continuum limit of a very large number of elements. In \S\,\ref{sec:LocalIC} we analyse the response of the system to a localized disturbance and show that a front propagates  through the system at constant speed; we examine the characteristics of this front propagation by determining its speed and the cluster size near the moving front. Finally, we examine the statistical distribution of cluster size from numerical simulations in \S\,\ref{sec:Stats} before summarizing our results and their implications in \S\,\ref{sec:Conclusions}.

\section{Theoretical Model}
\label{sec:model}

\begin{figure}
\centering
\includegraphics[width=0.75\textwidth,angle=0]{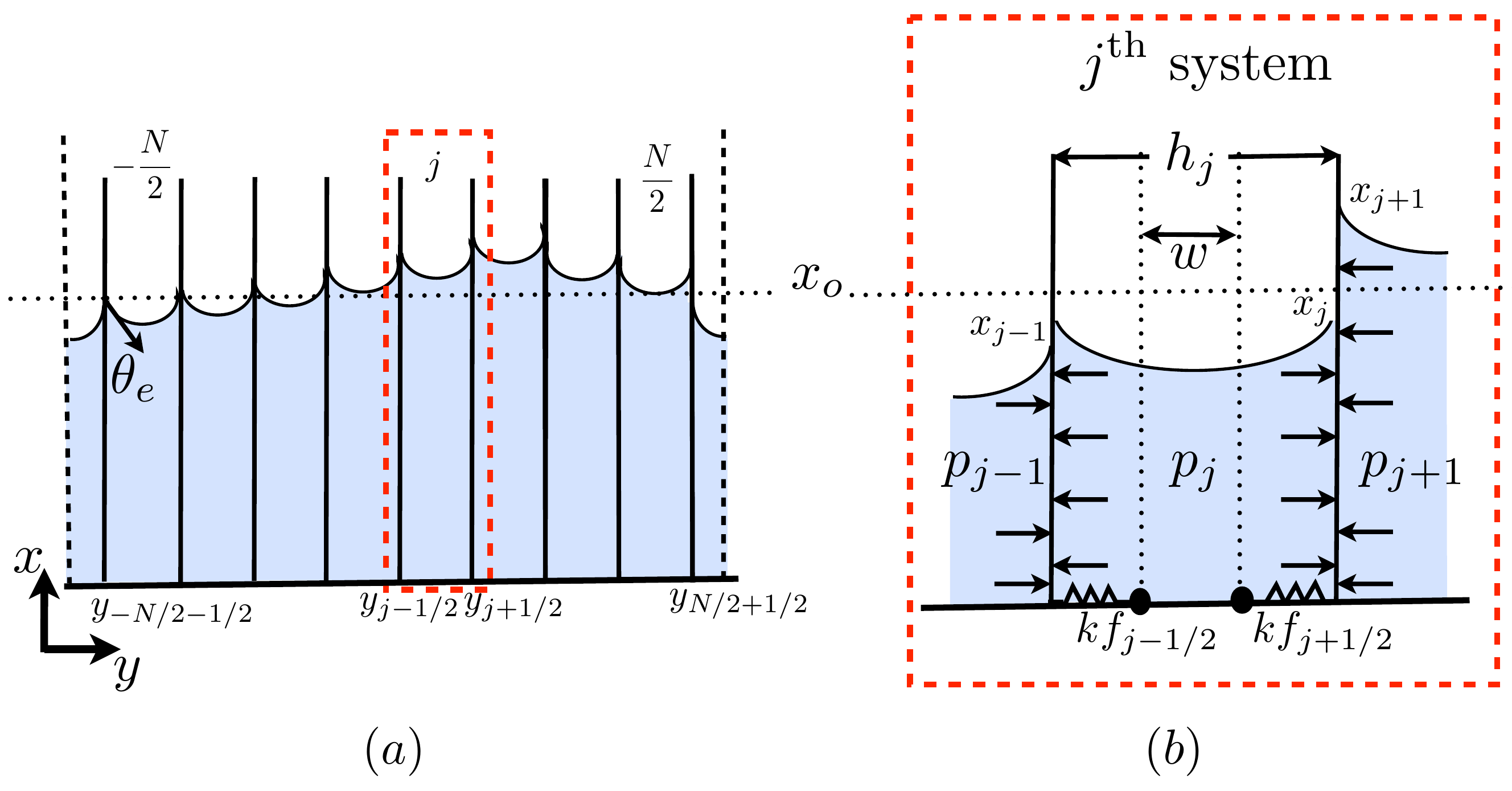}
\caption{(a) Sketch of a system of $N+2$ spring-block elements with liquid films filling the gaps between blocks. (b) A detailed view of two block-spring elements containing a liquid film. The initial configuration consists of equi-spaced blocks with inter-block gap, $w$, and liquid height $\xo$.}
\label{fig:sketch}
\end{figure}

We consider the interaction between multiple elastic elements induced by liquid placed in the gaps between them (see sketch in Figure~\ref{fig:sketch}). To isolate the physical effects of interest here (namely surface tension and elasticity), we neglect the effects of gravity and inertia (both of which are, in any case,  negligible at the small scales of MEMS applications); we assume there is no evaporation.

Models of elastocapillary dynamics typically treat the elastic elements as elastic beams with a bending stiffness, $B$, and length $L$ \cite[e.g.][]{Aristoff2011,Taroni2012}. The beams deflect due to a combination of surface-tension and viscous forces, which provide a spatially varying load on the beams. This treatment requires solution of a sixth-order parabolic partial-differential equation for the evolution of each beam shape. While studying  elastocapillary dynamics in such detail is tractable for two beams, here we wish to study the interaction of large numbers of elastic elements. We therefore simplify the problem by representing each elastic element by only a single displacement from its initial position and considering it to be subject to a restoring force proportional to this displacement; equivalently, we consider a linear array of rigid blocks that are each tethered to their initial position by a linear spring of stiffness $k$. Elementary beam theory shows that this simplification yields the correct load--displacement relationship for the cases of a point force or of a uniform pressure if the spring stiffness $k$ is chosen such that $k\propto B/L^3$, with the prefactor depending on the boundary conditions and loading \cite[see][]{Timoshenko1970}. This simplification of the problem has not, to our knowledge, been realized experimentally but is reminiscent of the elastocapillary instability of a helical elastic thread immersed in silicone oil \cite[][]{Jung2009}. The theoretical problem that we derive below is tractable and, as we shall see, contains a rich variety of phenomena.

\subsection{Governing equations}
\label{subsec:model}

Consider an array of $N+2$ rigid blocks that, when undisplaced, are uniformly spaced with a gap of width $w$ between neighbouring blocks.  Liquid of viscosity $\mu$ with surface tension $\gamma$ fills the gaps between the blocks to an initial height $\xo$ (see figure \ref{fig:sketch}b). We label the $N+1$ gaps between the blocks by integers $-N/2\leqslant j\leqslant N/2$ and  let the $(j+{1\over2})$th block have displacement, $f_{j+1/2}(t)$,  from its initial position. The block width does not affect the dynamics and, for simplicity, we assume it is negligible in comparison with $w$. Hence the positions of the blocks on either side of the $\ith$ gap are
\begin{align}
y_{j+1/2}&=\bigg(j+\frac{1}{2}\bigg)w+f_{j+1/2}(t), \notag\\
y_{j-1/2}&=\bigg(j-\frac{1}{2}\bigg)w+f_{j-1/2}(t),
\label{eqn:yi}
\end{align} and the width of the  gap is
\beq
h_{j}=y_{j+1/2}(t)-y_{j-1/2}(t) =w+f_{j+1/2}(t)-f_{j-1/2}(t).
\label{eqn:hdefn}
\eeq

Motion of the blocks causes liquid flow in the gaps. Using lubrication theory to model the flow, we find that the pressure in the $\ith$ gap, $p_j(x,t)$, satisfies the Reynolds equation
\begin{align}
\frac{h_j^3}{12 \mu }\frac{\partial^2p_j}{\partial x^2}=
\frac{\D h_j}{\dt}=\frac{\D f_{j+1/2}}{\dt}-\frac{\D f_{j-1/2}}{\dt}.
\label{eqn:reynoldseqn}
\end{align} 
We assume that the blocks sit on a rigid substrate at $x=0$ through which there is no fluid flux, so that
\beq
\left.\frac{\partial p_j}{\partial x}\right|_{x=0}=0.
\label{eqn:BC_pprime}
\eeq The meniscus in the $\ith$ gap is at height $x=x_j(t)$, where $x_j=x_0w/h_j(t)$ by mass conservation. Interfacial tension $\gamma$ acts on the meniscus curvature $\kappa$ to produce a pressure drop across the meniscus of magnitude $\gamma\kappa$. Taking the ambient pressure above the meniscus as  reference, and assuming the blocks to be perfectly wetting for simplicity (i.e.~$\theta_e=0$), we obtain
\beq
p_j(x_j,t)=-2\gamma /h_j.
\label{eqn:BC_p}
\eeq
(A non-zero contact angle simply replaces $\gamma$ by $\gamma\cos\theta_e$ in the non-dimensionalization if dynamic contact-angle effects can be neglected.)

The solution of \eqref{eqn:reynoldseqn} subject to the boundary conditions \eqref{eqn:BC_pprime} and \eqref{eqn:BC_p} is
\begin{equation}
p_j(x,t)= \frac{6 \mu }{h_j^3}\bigg(\frac{\D f_{j+1/2}}{\dt}-\frac{\D f_{j-1/2}}{\dt}\bigg)
\left(x^2-x_j^2\right)-\frac{2\gamma }{h_j},
\label{eqn:pressure}
\end{equation} and the total attractive force exerted on the  neighbouring blocks by the liquid in the $\ith$ gap is
\begin{equation}
F_j(t)= -\int_0^{x_j} p_j(x,t)\ \D x =\frac{4 \mu x_j^3}{h_j^3}\frac{\D h_j}{\dt}+\frac{2\gamma x_j}{h_j}.
\label{eqn:FP}
\end{equation} 
(A non-zero contact angle would add a force $\gamma\sin\theta_e$ to this equation, which is much smaller than the capillary pressure term.)  Each block is subject to a force of the form \eqref{eqn:FP} from the liquid columns on either side of it. Equating the resultant of these forces to the restoring force applied by the spring (since inertia has been neglected), we find that
\begin{equation}
kf_{j+1/2}=F_{j+1}-F_{j}.
\label{eqn:EOM1}
\end{equation} 

With $h_j$ and $x_j$ determined by \eqref{eqn:hdefn} and  mass conservation, and with suitable boundary conditions at the ends of the array, \eqref{eqn:FP} and \eqref{eqn:EOM1} form a system of first-order ordinary differential equations for the evolution of the block displacements $\{f_{j+1/2}(t)\}$. 

\subsection{Non-dimensionalization}

There are two characteristic and natural length scales in this problem:   the initial liquid  height, $\xo$, in the $x$-direction, and the initial gap width, $w$, in the $y$-direction. We use these to define a dimensionless gap thickness and meniscus position, $\bar{h}_j=h_j/w$ and $\bar{x}_j=x_j/\xo$, respectively. Conservation of mass in each gap thus reduces to $\bar{x}_j\bar{h}_j=1$.

A balance between the viscous lubrication pressure and the pressure jump due to surface tension in \eqref{eqn:pressure} yields a characteristic time scale $t_\gamma=2\mu\xo^2/\gamma w$. We define $\bar t=t/t_\gamma$ and, using mass conservation to eliminate $x_j$, obtain the dimensionless form of \eqref{eqn:FP} as
\begin{equation}
\bar{F}_j=\frac{F_j}{2\gamma \xo/w}=\frac{1}{\bar{h}_j^6}\frac{\D\bar{h}_j}{\D\bar{t}}+\frac{1}{\bar{h}_j^2}, 
\label{eqn:ND_FP}
\end{equation} 
Note that the time scale $t_\gamma$ introduced here differs from the elastocapillary time scale, $\mu (B/\gamma^3)^{1/2}$, that is usually used \cite[e.g.][]{Taroni2012}. In particular, $t_\gamma$ is independent of the elasticity in the system and is, instead, a characteristic time scale for capillary pressure to suck liquid to a height $\xo$ in a channel of width $w$.

Non-dimensionalizing the force balance  \eqref{eqn:EOM1}, we obtain
\beq
2\kec \bar{f}_{j+1/2}=\bar{F}_{j+1}-\bar{F}_{j},
\label{eqn:EOM1ndim}
\eeq where
\beq
\kec=\frac{kw^2}{4 \gamma \xo}
\label{eqn:kec}
\eeq is the dimensionless spring stiffness, which compares the spring force produced by displacement of the block by a distance $w$ with the total force due to capillary suction $2\gamma\xo/w$ in the equilibrium state. The dimensionless parameter $\kec$ is analogous to the \emph{elastocapillary number} introduced by \cite{Mastrangelo1993}, modulo some differences due to the different geometry considered there. Henceforth we drop the bars on all dimensionless variables.

The non-dimensionalization presented above shows that the behaviour of the system is governed by a single dimensionless parameter, the spring stiffness $\kec$. To obtain some physical understanding of the important role played by $\kec$, we first consider the predictions of our model for an isolated pair of blocks and then present illustrative numerical results for a large number of interacting blocks.

\subsection{An isolated pair}
\label{subsec:pairs}

The simplest problem to consider is that of two isolated blocks with liquid between them (but none outside). This problem is similar to that considered by \cite{Taroni2012} for two elastic beams clamped at one end and free at the other. With no fluid outside the blocks, $F_1=F_{-1}=0$, and thus there is symmetry about the mid-plane with $f_{1/2}=-f_{-1/2}=-F_0/(2\kec)$. To derive the evolution equation for the gap $h(t)$ between the blocks, we note that $h=1+f_{1/2}-f_{-1/2}=1-F_0/\kec$. From \eqref{eqn:ND_FP} we deduce 
\beq
\frac{1}{h^6}
\frac{\D h}{\dt}=-\frac{1}{h^2}+ \kec(1-h).
\label{eqn:EOMpair}
\eeq

\begin{figure}
\begin{center}
\mbox{
\begin{tabular}{cc}
\subfigure[]
{\includegraphics[width=0.45\textwidth,angle=0]{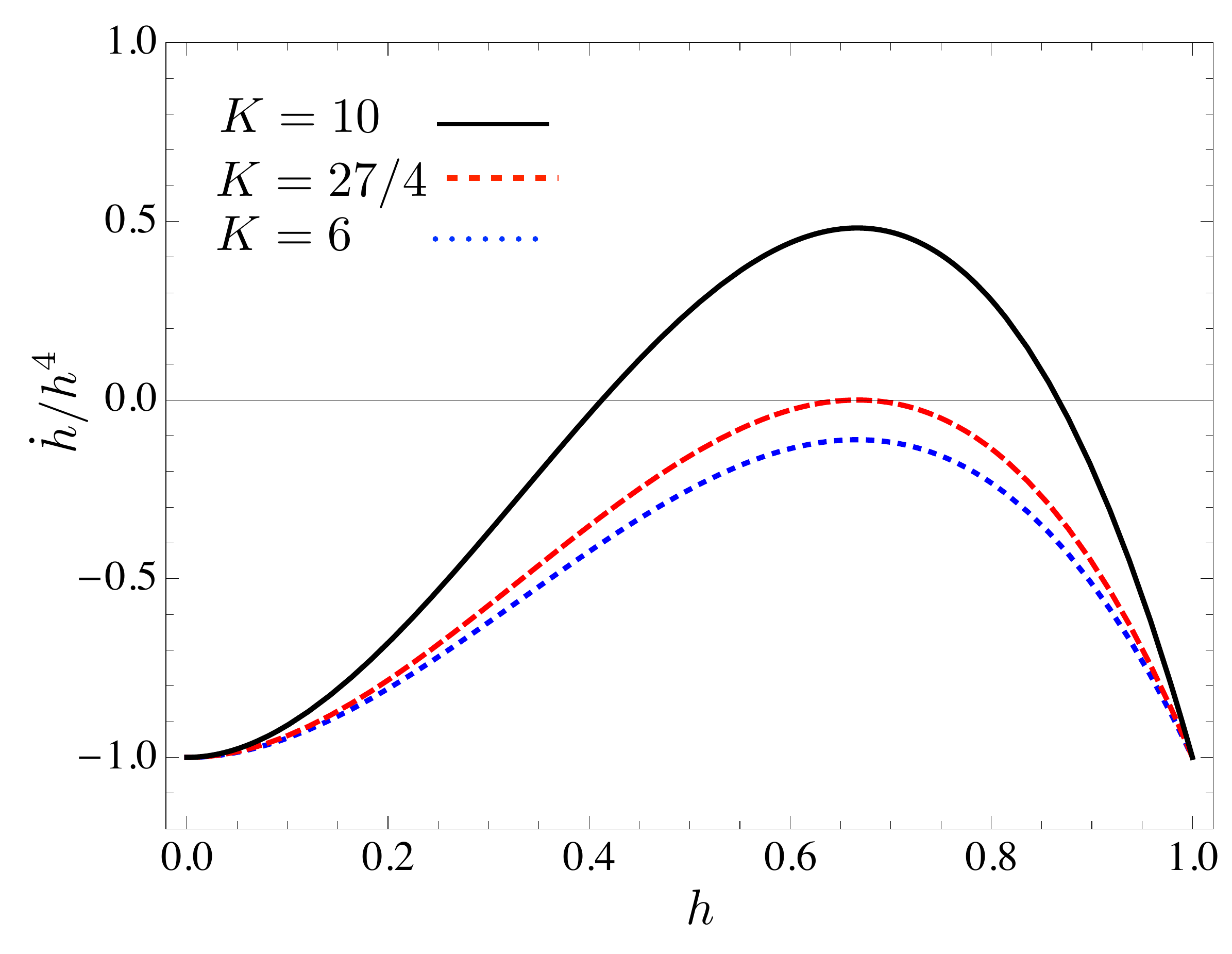}
}
&
\subfigure[]
{\includegraphics[width=0.45\textwidth,angle=0]{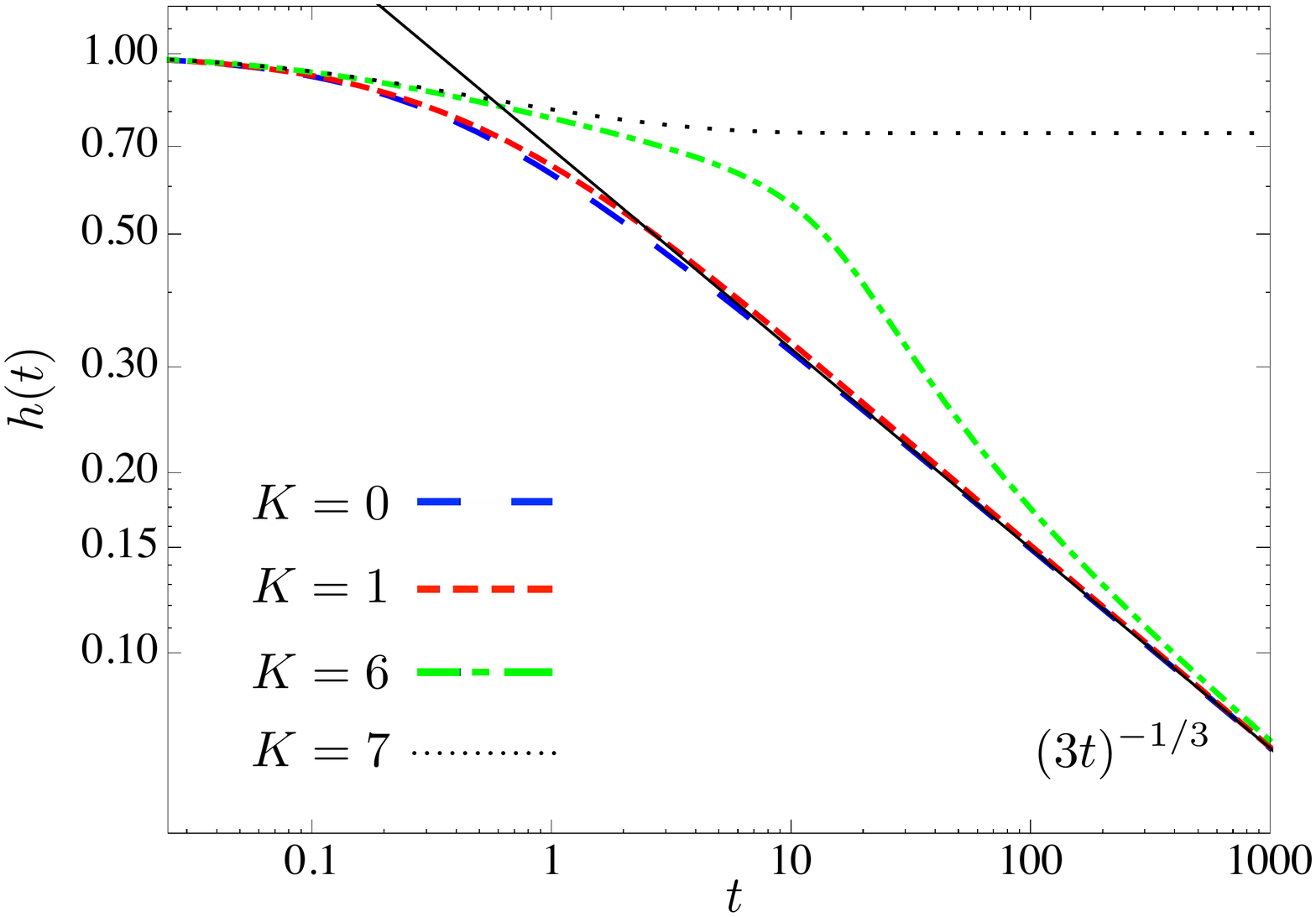}
} 
\end{tabular}
}
\end{center}
\caption{Dynamics of an isolated pair: (a) The phase-plane plot shows that non-zero equilibrium solutions exist for $\kec\geq 27/4$. 
(b) For $\kec<27/4$ (i.e.~with no separation at equilibrium) numerical solutions (solid curves) tend to the universal late-time behaviour given in \eqref{eqn:h_asyp} (black dashed line). For $\kec>27/4$ a non-trivial equilibrium separation is attained (dotted curve).}
\label{fig:2dyn}
\end{figure}

The first term on the right-hand side of \eqref{eqn:EOMpair} represents the capillary attraction between the blocks due to the negative capillary pressure below the meniscus, while the second term represents the restoring effect of the springs. If $\kec=0$ (no springs) then the gap decreases monotonically towards $h=0$ (contact between the blocks), but does so ever more slowly due to the increasing viscous resistance in the closing gap, which increases even more rapidly (like $h^{-6}$) than the capillary attraction ($h^{-2}$). 

If $\kec>0$ then the blocks still start to be attracted together with $\D h/\D t=-1$ from the initial separation $h(0)=1$, but there is a possibility that, as the restoring force from the springs increases from zero, the gap spacing can approach a nonzero equilibrium. Nonzero equilibria correspond to any positive real roots of the  cubic $\kec(1-h)h^2=1$. (The negative real root corresponds to a negative gap thickness, which is unphysical.) Consideration of the variation with $\kec$ (see figure \ref{fig:2dyn}{\it a}) shows that if $0<\kec<27/4$ then there are no such equilibria, and the gap thickness decreases monotonically towards zero. For $\kec>27/4$ there are two equilibria in the range $0<h<1$; from the sign variations of $\D h/\dt$, the larger one is stable and the smaller one is unstable. 

The contacting equilibrium, $h=0$, is always stable. In particular, examining the behaviour for $h\ll1$ shows that, as $t\to \infty$,
\beq
h\sim (3t)^{-1/3}+K/(3t)
\label{eqn:h_asyp}
\eeq
which is independent of $\kec$ at leading order (figure \ref{fig:2dyn}{\it b}).
The contacting equilibrium is therefore approached only algebraically slowly, governed by a dominant capillary--lubrication balance. If $\kec>27/4$ and there is another stable equilibrium, which of the two stable equilibria is approached at late times depends on the initial condition.

These results accord qualitatively with the more detailed calculations of \cite{Taroni2012} for the case of two elastic beams. In particular, if too much liquid is placed between the beams or the beams are not stiff enough ($\xo$ too large, or $k$ too small, in our notation) then the only equilibrium is in contact; otherwise multiple equilibria are possible and the late-time behaviour is dependent on initial conditions. In our model non-contacting equilibrium solutions only exist if the dimensionless spring stiffness $\kec$ is large enough to balance the capillary suction, and for relatively weak springs ($\kec<27/4$ in the two-block problem) surface tension acts to bring the blocks into contact. Contact is approached algebraically slowly as the viscous resistance in the closing gap increases rapidly. We now investigate how these ideas are modified as we return to our main focus --- the interaction of many elastic elements.

\subsection{Many elastic elements}
\label{subsec:multieqns}

Consider again a large array of $N+2$ blocks. Away from the ends, 
substitution of the dimensionless force \eqref{eqn:ND_FP} into the force-balance equation \eqref{eqn:EOM1ndim} yields
\beq
2 \kec f_{j+1/2}=-\frac{1}{h_j^6}\frac{\D h_j}{\dt}+\frac{1}{h_{j+1}^6}\frac{\D h_{j+1}}{\dt}
-\frac{1}{h_j^2}+\frac{1}{h_{j+1}^2}.
\label{eqn:NEOM1}
\eeq
for the $(j{+}\frac{1}{2})$th block. The dimensionless gap thickness is given by $h_j=1+f_{j+1/2}-f_{j-1/2}$ from \eqref{eqn:hdefn}. Combining this with \eqref{eqn:NEOM1}, we obtain an evolution equation for the gap thicknesses  alone, namely
\begin{align}
2 \kec(h_j-1)=\frac{1}{h_{j+1}^6}\frac{\D h_{j+1}}{\dt}-\frac{2}{h_j^6}\frac{\D h_j}{\dt}+\frac{1}{h_{j-1}^6}\frac{\D h_{j-1}}{\dt}+\bigg(\frac{1}{h_{j+1}^2}-\frac{2}{h_{j}^2}+\frac{1}{h_{j-1}^2}\bigg).
\label{eqn:NEOM2}
\end{align}

At the ends, we must prescribe conditions on the first and last blocks, such as zero displacement, zero external force or periodicity. In the simulations presented below, we applied symmetry boundary conditions on $h$ by introducing `ghost' points outside the array with $h_{-N/2-1}=h_{-N/2+1}$ and $h_{N/2+1}=h_{N/2-1}$.  Equation \eqref{eqn:NEOM2} then applies for $-N/2\leqslant j\leqslant N/2$ and provides a system of $N+1$ first-order ordinary differential equations for the $h_j$.  Note that with these boundary conditions (or with zero-displacement or periodic conditions), and contrary to the case of an isolated pair, the undeformed state $h_j=1$ is an equilibrium solution of \eqref{eqn:NEOM2}.

\subsubsection{Numerical experiments}

The behaviour of this multi-block system mirrors that observed for an isolated pair in the respect that interactions between blocks initially occur on an $O(1)$ time scale but, as some of the gaps between blocks get smaller, lubrication forces greatly slow down their dynamics. In a large system with many such interactions and an increasing spread of gap sizes, a range of different time scales is thus at work and the system of equations becomes stiff. The differential system can, nevertheless, be solved efficiently and accurately using adaptive time-stepping and a semi-implicit scheme with linear terms treated implicitly and non-linear terms explicitly \cite*[see][for a similar idea]{Lister2010}.  

\begin{figure}
\centering
\includegraphics[width=1.01\textwidth,angle=0]{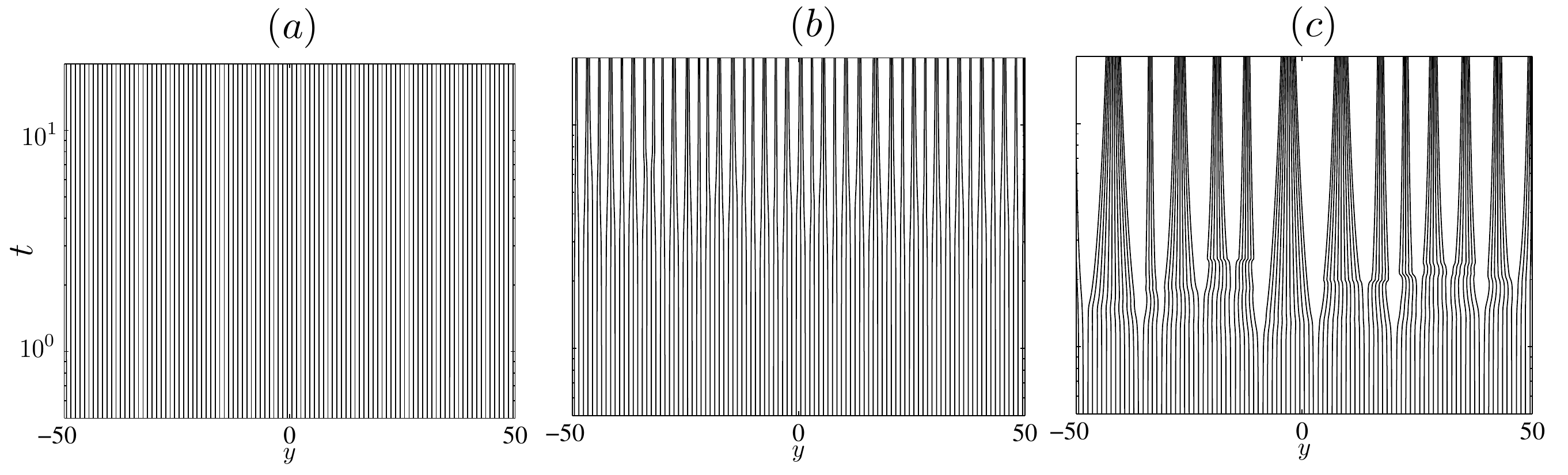}
\caption{Spatio-temporal plots of the evolution of an array of blocks initially subject to a random perturbation of the form \eqref{eqn:ran_ic} with $\epsilon=10^{-2}$. Here $\Ntot=100$ and results are shown for three different spring stiffnesses: (a) $\kec=5$, (b) $\kec=1$, and (c) $\kec=0.1$. Each curve in $(y,t)$ space corresponds to the trajectory of a single block.
} 
\label{fig:nums_ran}
\end{figure}

To illustrate the typical behaviour of the system, figure \ref{fig:nums_ran} shows spatio-temporal plots for the evolution of the block positions with three different values of the spring stiffness $\kec$. In each case, the initial condition was a small, spatially random perturbation to the equilibrium state,
\begin{equation}
h_j(t=0)=1+\epsilon \mathcal{R}_j,
\label{eqn:ran_ic}
\end{equation}
where $\mathcal{R}_j$ is a  random number sampled from a uniform distribution on $[0,1]$ and $\epsilon\ll1$.

For sufficiently large values of $\kec$ (stiff springs) very little happens: the perturbed state decays towards the equilibrium state $h_j=1$ and the blocks remain isolated (figure \ref{fig:nums_ran}{\it a}). However, for small values of $\kec$ (compliant springs)  the blocks cluster into groups of different sizes with the typical size of these clusters apparently increasing with decreasing $\kec$ (figure \ref{fig:nums_ran}{\it b,c}). Both of these observations tie in with the qualitative behaviour seen in figure \ref{fig:pics}: large values of $\kec$ correspond to very weak surface tension, as in figure \ref{fig:pics}({\it a}), where no deformation was observed. On the other hand, small aspect ratio corresponds to small $\kec$ (since for an elastic beam of thickness $h$ and length $L$, $k\sim (h/L)^3$); the result that larger clusters are formed with small $\kec$ therefore ties in with the observation in figure \ref{fig:pics}({\it c}) that small-aspect-ratio beams form large clusters. Finally, we note the apparent `kinks' in the spatio-temporal plot for $\kec=0.1$ (figure \ref{fig:nums_ran}{\it c}); these arise due to coalescence or separation events between clusters, which occur on a much shorter time scale than the slow collapsing of a single cluster once it has formed. The final range of cluster sizes is greater for $\kec=0.1$ than for $\kec=1$. The cluster-size distribution is investigated in more detail in \S\,\ref{sec:Stats}.

\subsubsection{Collapse of a cluster} 

The numerical results in figure \ref{fig:nums_ran}({\it c}) show clusters becoming isolated relatively quickly, but continuing to evolve by collapsing on a slower time scale. To understand the collapse of a single cluster, we  perform a late-time asymptotic analysis of the system in Appendix~\ref{app:asymp_clump}. For the case that all clusters are of size $M$, this analysis shows that the width of the $m^{\mathrm{th}}$ gap (counting from one end) evolves as 
\beq
h_{m}\sim (3t)^{-{1/3}}+\big\{\kec m(M-m)+M^{-2}\big\}/(3t),
\label{eqn:hl_asymp}
\eeq
where the term in curly brackets is the force $F_m$ in the gap. At leading order, we see that all gaps decay algebraically like $(3t)^{-1/3}$ in just the same way as an isolated pair, cf.~\eqref{eqn:h_asyp}, due to the same dominant capillary--lubrication balance. The $O(t^{-1})$ correction to this leading-order behaviour has a parabolic dependence on the distance from the centre of the cluster, which may be understood physically as follows. The capillary attraction collapsing the outermost gap is opposed by the capillary attraction $M^{-2}$ to the neighbouring cluster and the spring force on the outermost block, the capillary attraction of the next gap inward is additionally opposed by the spring force on the next block inward, and the capillary attraction in the central gap is opposed by the effect of all the spring forces in the cluster which pull to left and right to slow the closure of the central gap. (See Appendix~\ref{app:asymp_clump} for details.) When the gaps are small, the capillary attraction is dominant, but the combined effect of the springs and the attraction to neighbouring clusters is to slow the rate of collapse a little (more so than for an isolated pair); the slowing due to the spring forces is more pronounced in the middle of a cluster than at the edges.

\subsubsection{A continuum approximation}

The numerical results presented in figure \ref{fig:nums_ran}  suggest that for $\kec\ll1$ the system evolves into a series of clusters, each containing a large number of blocks. To describe such long-wavelength structures, it is instructive to consider the governing equation \eqref{eqn:NEOM2} as the natural discretization of a partial-differential equation. Using  the mapping
\begin{align}
(.)_{j+1}-(.)_{j}\to \frac{\partial (.) }{\partial y},\ \quad
(.)_{j+1}-2(.)_{j}+(.)_{j-1} \to\frac{\partial^2(.) }{\partial y^2}, 
\label{eqn:approx}
\end{align} 
there is an obvious correspondence to the partial-differential equation 
\beq
\frac{\partial^2}{\partial y^2}\bigg(\frac{1}{h^6}\frac{\partial h}{\partial t}
+\frac{1}{h^2} \bigg)=2 \kec(h-1)~.
\label{eqn:EOM2_Cont}
\eeq 
Here $y$ is a continuous Lagrangian coordinate, $-l/2\leqslant y\leqslant l/2$, analogous to the discrete coordinate $-N/2\leqslant j \leqslant N/2$, and $l$ is the continuum domain length. We shall use \eqref{eqn:EOM2_Cont} to complement our analysis of the discrete system of ordinary differential equations in what follows.

\section{Linear stability analysis}
\label{sec:linstab}

The numerical simulations presented in \S\,\ref{subsec:multieqns} demonstrate that, for sufficiently small $\kec$, the undeformed equilibrium state $h_j=1$ is unstable to small perturbations. We examine the linear stability of the undeformed state explicitly for both the discrete and continuum models.

\subsection{Discrete problem\label{sec:discretelinstab}}

Consider a small, periodic disturbance to the undeformed state of the form $h_j=1+\epsilon\exp(2\pi\jj j/\Np+\sigma t)$ with $\epsilon\ll 1$, where $\Np$ is the number of blocks in each period of the disturbance and $\jj=\sqrt{-1}$. Substituting this ansatz  into \eqref{eqn:NEOM2} and linearizing, we find that the growth rate of the perturbation is given by
\begin{equation}
\sigma=2-\frac{\kec}{2\sin^2(\pi/\Np)}.
\label{eqn:sigd}
\end{equation} From this expression, we conclude that for $\kec<4$ the perturbation is unstable if
\beq
\Np<\Npmax(\kec)=\frac{\pi}{\sin^{-1}[(\kec/4)^{1/2}]},
\label{eqn:discretemaxN}
\eeq and that, independently of $\kec$, the most unstable (or least stable) mode is given by $\Np=2$. (\cite{Gat2013} also found that pairwise clustering is the fastest growing linear-instability mode in their model.) Finally, we observe that for $\kec>4$ any periodic perturbation decays ($\sigma<0$ for all $\Np$), and we conclude that the undeformed state is stable for $\kec>4$. 

These results explain some features of the numerical results presented in figure \ref{fig:nums_ran}. For example, the prediction that the undeformed state is stable to small perturbations for $\kec>4$ is consistent with figure \ref{fig:nums_ran}({\it a}), which showed stability for $\kec=5$. Furthermore, \eqref{eqn:discretemaxN} suggests that the system is only unstable to clusters below a maximum size and that this maximum increases with $\kec$. This is in qualitative accordance with the results seen in figures \ref{fig:nums_ran}({\it b}) and \ref{fig:nums_ran}({\it c}).

However, \eqref{eqn:sigd} shows that the fastest growing mode has $\Np=2$, whereas our numerical results show that, in fact, a range of cluster sizes are seen. Moreover, the clustering in the numerical experiments does not evidently even begin pairwise, but rather seems to have a range of cluster sizes as soon as they emerge from the initial random perturbation. This is not necessarily a contradiction since \eqref{eqn:sigd} shows that for $\kec\ll1$ and $\Np\ll\Npmax(\kec)$ the growth rate is relatively insensitive to the precise value of $\Np$ --- in this case several sizes of cluster might have much the same growth rate and hence be observed experimentally. Nevertheless, the contrast between the fastest growing mode and the observed range of cluster sizes is somewhat of a surprise, and we will argue in \S\,\ref{sec:LocalIC} that the characteristic cluster size emerges from quite different considerations.

\subsection{Continuous problem}

Consider a small, periodic disturbance of the form $h=1+\epsilon\exp(\jj \ell y+\sigma t)$ in the partial-differential equation \eqref{eqn:EOM2_Cont}. Upon linearizing, we find that the growth rate of periodic perturbations in the continuous problem is given by
\begin{equation}
\sigma=2\bigg(1-\frac{\kec}{\ell^2}\bigg),
\label{eqn:sigc}
\end{equation} 
where $\ell$ is the wavenumber.  Note that discrete perturbations with period $\Np$ are analogous to continuous perturbations with wavenumber $\ell_p=2\pi/\Np$, and that in the limit $\Np\gg1$ \eqref{eqn:sigd} reduces to \eqref{eqn:sigc} with $\ell\to\ell_p$. Hence the discrete growth rate reproduces the continuous growth rate for large-wavelength disturbances, as expected.

Equation \eqref{eqn:sigc} shows that continuous perturbations are stable whenever the wavenumber $\ell>\kec^{1/2}$. This suggests a maximum cluster length $N_{\mathrm{max}}=2\pi \kec^{-1/2}$, which is in agreement with the result \eqref{eqn:discretemaxN} of the discrete analysis in the limit $\kec\ll1$. Furthermore, the fastest growing instability in the continuum problem is that with the largest possible wavenumber $\ell$, or smallest possible wavelength, as was also found for the discrete problem. 

\section{Localized perturbations}
\label{sec:LocalIC}

Having considered the linearized response of an array of blocks to a global, periodic perturbation, we now consider the opposite limit of a small perturbation initially localized to just one block. To give a sense of the response to such a localized initial perturbation, we first present sample results of numerical simulations. We then analyse this problem in detail.

\subsection{Numerical simulations}
\label{subsec:numerics_wave}

\begin{figure}
\centering
\includegraphics[width=0.995\textwidth,angle=0]{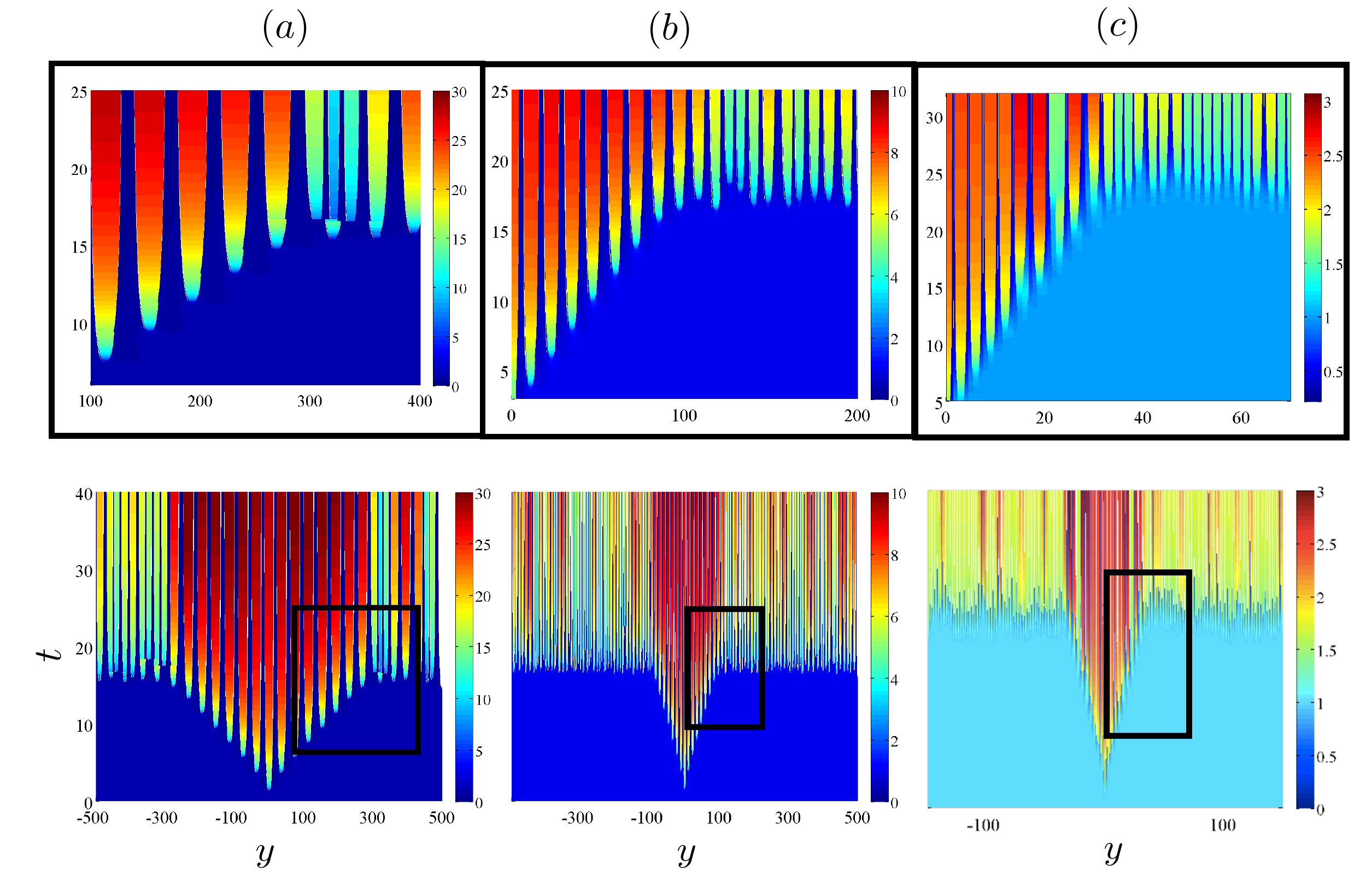}
\caption{Spatio-temporal diagram showing the evolution of the gap-width $h_j(t)$ (indicated by colour level) as a function of space and time for an initial condition  localized about $y=0$ for the cases ({\it a}) $\kec=0.01$, ({\it b}) $\kec=0.1$, ({\it c}) $\kec=1.0$. Observe the propagation of the disturbance away from its initial location at $y=0$, the formation of mono-disperse clusters in the vicinity of the front and of poly-disperse clusters growing from noise in the far-field. Here the initial condition is of the form \eqref{eqn:IC_delta} with $\epsilon=0.01$ and $\Ntot=10^3$.}
\label{fig:wave}
\end{figure}

Numerical simulations were performed for a range of values of $\kec$ with the initial condition
\begin{equation}
h_j(t=0)=1+\epsilon\delta_{j,0},
\label{eqn:IC_delta}
\end{equation}  
where $\epsilon\ll1$ and $\delta_{j,k}$ is the Kronecker delta. The results  presented in figure \ref{fig:wave} show that the effect of the localized perturbation propagates away from $j=0$ and ultimately leads to clumping throughout the system. The spatio-temporal diagrams show that the disturbance propagates at a roughly constant speed into the undisturbed region (hence the wedge-shaped region of disturbance), though after a certain time the inherent instability of the system takes over and gives global growth of disturbances. Two features of these solutions are of particular interest here: Firstly the propagation speed of the local disturbance appears to decrease with increasing  $\kec$. Secondly, the zoom-in insets of figure \ref{fig:wave} reveal that, as the wedge-shaped front passes a given location, monochromatic clusters are formed with a cluster size that decreases as $\kec$ increases. We shall explain both of these features by detailed analysis in 
\S\,\ref{sec:LinearProp}. A third feature of these solutions, that the inherent global instability becomes evident after a time that increases with increasing $\kec$, can be understood qualitatively by the decrease in the maximum linear growth rate \eqref{eqn:sigd} with increasing $\kec$.

\subsection{Linearized analysis}\label{sec:LinearProp}

The initial perturbation \eqref{eqn:IC_delta} applied in the numerical simulations shown in figure \ref{fig:wave} is small. It is therefore natural to seek a solution of the form $h_j(t)=1+\epsilon H_j(t)$, where $\epsilon\ll1$, and linearization of \eqref{eqn:NEOM2} then yields
\begin{align}
\frac{\D H_{j+1}}{\dt}-{2}\frac{\D H_j}{\dt}+\frac{\D H_{j-1}}{\dt}-2\bigg({H_{j+1}}-{2H_{j}}+{H_{j-1}}\bigg)=2 \kec{H_j},
\label{eqn:LinH}
\end{align} 
to be solved subject to the initial condition
\beq
H_j(t=0)=\delta_{j,0}.
\label{eqn:LinHIC}
\eeq 

\begin{figure}
\centering
\includegraphics[height=0.45\textwidth,angle=0]{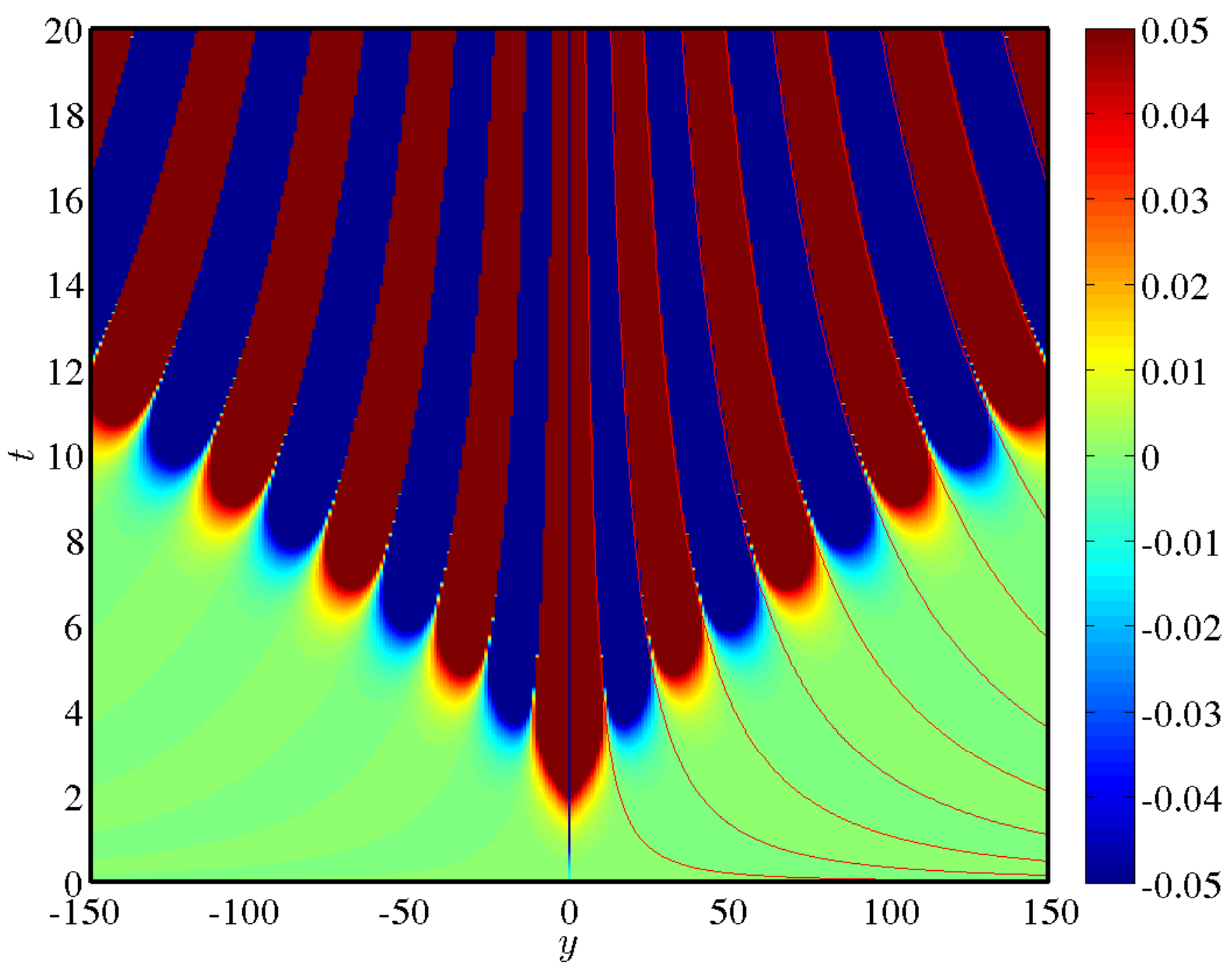}
\caption{Spatio-temporal plot showing the evolution of the disturbance to the gap width $H_j(t)$ from a localized perturbation in the linearized problem. On the right-hand side, curves of constant $y^2t$ are plotted according to the phase in \eqref{eqn:statphasecont}; the value of $\psi_0$ is chosen so that these curves pass through the zeros of $H$. Here $\kec=0.01$, $\epsilon=0.01$ and $N=10^3$.}
\label{fig:LinNums}
\end{figure}

Before proceeding  further with the analysis, it is as well to check that the dynamics of the linearized system is qualitatively similar to that of the fully nonlinear system. To do this, we solved \eqref{eqn:LinH} subject to the initial condition \eqref{eqn:LinHIC}, and we present the resultant  spatio-temporal plot in Figure~\ref{fig:LinNums}. The response to a localized perturbation in the linearized problem -- constant-speed invasion of the undisturbed region, forming monochromatic clusters -- is indeed qualitatively similar to that in the fully nonlinear problem (presented in figure \ref{fig:wave}{\it a}).  (A minor difference is that there is no saturation of amplitudes at late times, owing to the lack of the $h^{-6}$ factors in the lubrication resistance of the fully nonlinear problem; the colour scale in figure \ref{eqn:LinHIC} has therefore been chosen so that the continued unbounded growth does not obscure the first appearance of clusters.)

\subsubsection{Continuous problem\label{sec:contlocalized}}

Although our main interest lies in the discrete problem, a linearized analysis of the continuous problem yields important insights into the behaviour of the system. We therefore consider the continuum limit of the linearized equation \eqref{eqn:LinHIC}, which reads
\begin{equation}
\frac{\partial^3H}{\partial y^2\partial t}-2\frac{\partial^2H}{\partial y^2}=2\kec H.
\label{eqn:pde_H}
\end{equation} This may be simplified by letting $H(y,t)=e^{2t}\eta(y,t)$ and then introducing a rescaled time variable $\tau=2\kec t$ so that
\begin{equation}
\frac{\partial^3\eta}{\partial \tau\partial y^2}=\eta.
\label{eqn:pde_f}
\end{equation}
Using Fourier transforms, the solution of \eqref{eqn:pde_f} is
\begin{equation}
\eta(y,\tau)=\frac{1}{2\pi}\int_{-\infty}^{\infty}\zh_0(\ell)\exp(\jj \ell y- \tau/\ell^2)~\upd \ell,
\label{eqn:f_st}
\end{equation}
where $\zh_0(\ell)$ is the Fourier transform of some initial condition $\eta(y,0)$. 

To understand the behaviour of the solution in \eqref{eqn:f_st}, we let $y=c\tau$ and use the method of steepest descents \cite[e.g.][]{Hinch1990} to examine the behaviour in the limit $\tau\gg1$. After some standard analysis, we find that 
\beq
H(y,t)\approx I_0\frac{2^{5/6} \kec^{1/6}}{3^{1/2}\pi^{1/2}}\frac{t^{1/6}}{y^{2/3}}\exp\bigg[2t\bigg(1-\frac{3\kec^{1/3}}{2^{7/3}}\frac{y^{2/3}}{t^{2/3}}\bigg)\bigg]
\cos\bigg[ \frac{3^{3/2}}{2^{4/3}}(\kec y^{2}t)^{1/3}- \psi_0\bigg],
\label{eqn:statphasecont}
\eeq
where $I_0, \psi_0$ are the amplitude and phase of $\zh_0$ at the relevant saddle points (see Appendix~\ref{app:SP_C} for more details). The most important feature of the solution \eqref{eqn:statphasecont} is that, along a given ray $y/t=\mathrm{const.}$, the displacement $H(y,t)$  grows with time if
\begin{equation}
\frac{y}{t}\leqslant \cmax=\frac{2^{7/2}}{3^{3/2}}\kec^{-1/2}\approx 2.18 \kec^{-1/2}.
\label{eqn:ybyt}
\end{equation}
This result explains why growing disturbances are confined to a wedge in the $(y,t)$-plane, and predicts the `wavespeed' that is seen in simulations (see Figure~\ref{fig:waves_speeds}).

Another feature of the solution  \eqref{eqn:statphasecont} is the oscillatory behaviour in the cosine term. In this linearized solution the wave crests propagate towards the origin as $t$ increases (figure \ref{fig:LinNums}). In the full nonlinear problem (figure \ref{fig:wave}), disturbances have grown to be $O(1)$ within the wedge, and we do not expect, or observe, continued propagation of crests as the waves saturate quickly and $h\to0$ between the clusters. Nevertheless, it is reasonable to assume that the linearized analysis  is a good approximation of the nonlinear problem outside and up to the leading edge of the disturbance wedge where the disturbance is still small.  If we hypothesize that the wavelength of the saturated nonlinear clusters is set, or frozen in, near the leading edge of the wedge, then we would expect that the cluster size $\Np$ should be given by the frontal wave speed $\cmax$ divided by the frequency of oscillation in \eqref{eqn:statphasecont} along the line $y=\cmax t$. This calculation yields
\beq
\Np\approx \frac{2^{7/2}\pi}{9}\kec^{-1/2}.
\label{eqn:Nmax:cont}
\eeq 
and gives excellent agreement for $\kec\ll1$ with the nonlinear results (see figure \ref{fig:waves_speeds}\emph{b}). 

\subsubsection{The discrete problem\label{sec:discretelocalized}}

Although similar in principle to the above analysis of the continuous system, there are several subtleties that make the discrete calculation more intricate --- these details are left for Appendix \ref{app:SP_D}. In summary, we use similar  rescalings as for the continuous problem and take a Laplace transform in time to obtain a set of difference equations for the Laplace transforms of $\{H_je^{-2t}\}$, which can be solved analytically. Applying an inverse Laplace transform and using the method of steepest descents again, we find that
\beq
H_j(t)\sim \sum_{k}
A_k\exp[(2- \gjk\kec)t ] ,
\label{eqn:Lin_D_H}
\eeq
where
\beq
\gjk= \frac{2\jj j}{\kec t}~\ths_{j,k}+\frac{1}{2\sin^2\ths_{j,k}}
\eeq and the sum in \eqref{eqn:Lin_D_H} is over the relevant subset of the roots $\{\ths_{j,k}\}$ of the equation
\begin{align}
 \frac{2\jj j}{\kec t}\tan^3\theta-\tan^2\theta-1=0
\label{eqn:Lin_D_SP}
\end{align} that arises from finding saddle points in a steepest-descents calculation.

\begin{figure}
\centering
\mbox{
\begin{tabular}{cc}
\subfigure[]
{\includegraphics[width=0.475\textwidth,angle=0]{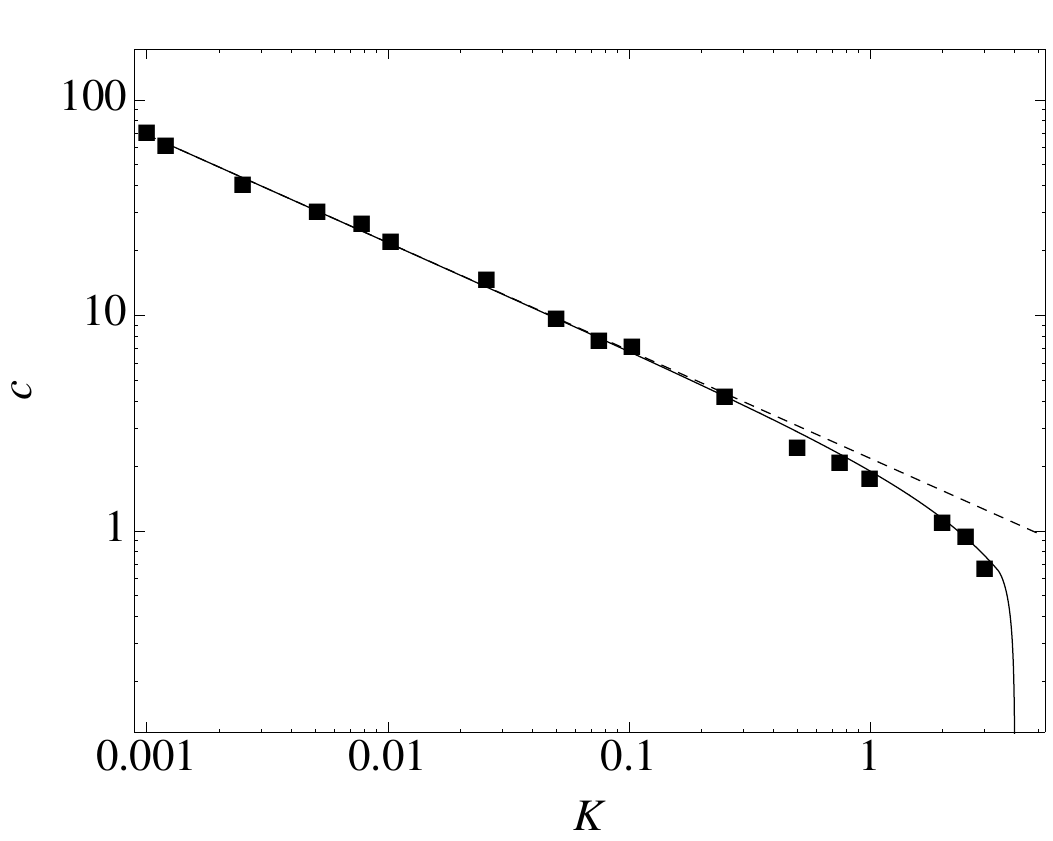}
}
&
\subfigure[]
{\includegraphics[width=0.475\textwidth,angle=0]{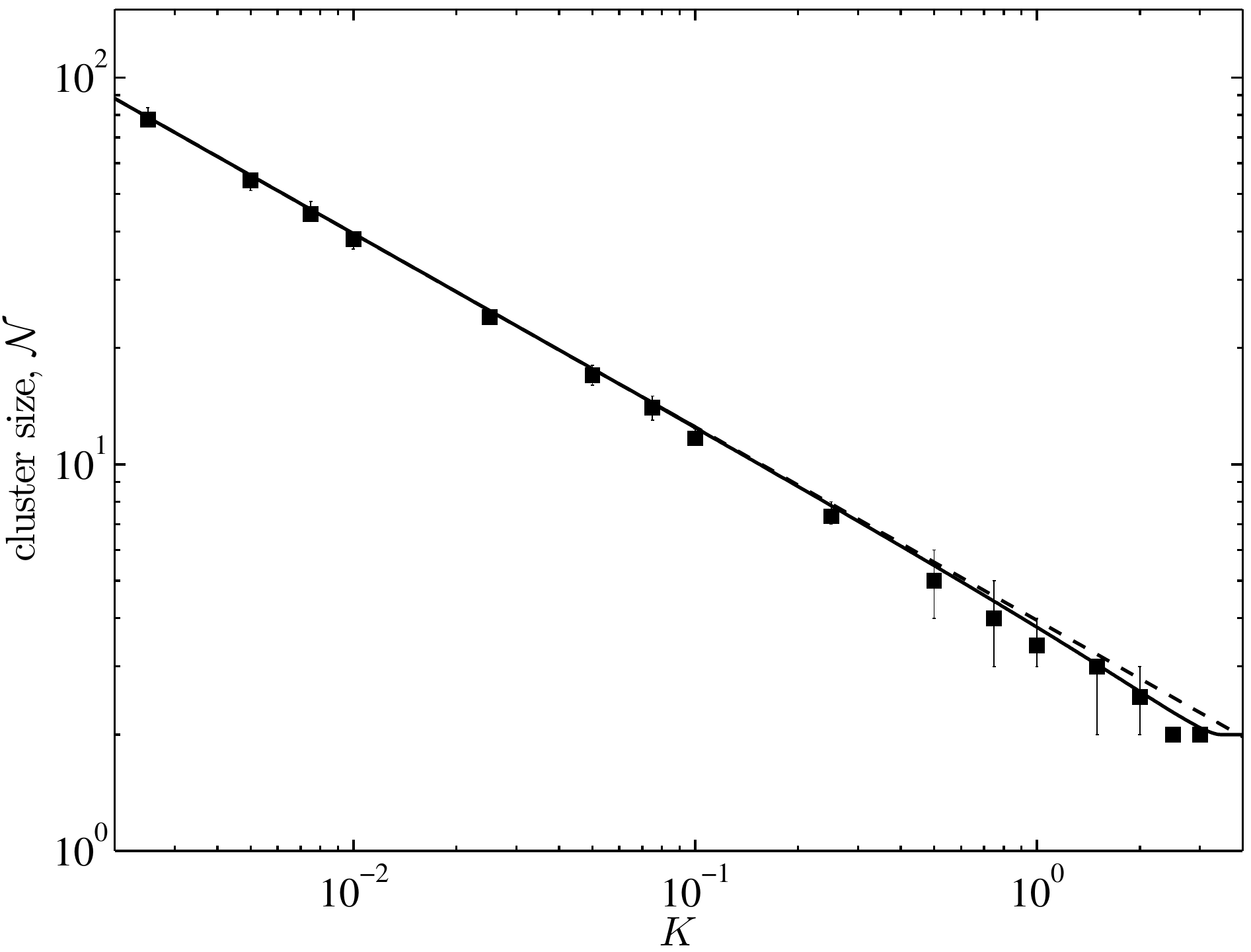}
}
\end{tabular}
}
\caption{ Numerical results from calculations of propagation from a localized disturbance compared to asymptotic predictions from the linearized problem. The dependence of ({\it a})  the speed of propagation of the front and ({\it b}) the cluster size observed at the front on the spring stiffness, $\kec$. Numerical results ($\blacksquare$) were determined from a computation on a  grid moving with the wavefront. These results compare well with the predictions of the discrete analysis (solid curves), and, for $\kec\ll 1$, the continuous analysis \eqref{eqn:ybyt} (dashed line). The vertical error bars in (\emph{b}) correspond to the observed range of cluster sizes with the mean value plotted as $\blacksquare$.}
\label{fig:waves_speeds}
\end{figure}

As in the continuous case, the important feature of the solution \eqref{eqn:Lin_D_H} is that, along a given ray, $j/t=\mathrm{const}$, the disturbance will grow with time if
\beq
\Real\{\gjk\} < 2/\kec.
\label{eqn:discreteineq}
\eeq For a given value of $\ct=j/\kec t$, it is a simple matter to calculate the values of $\gjk(\ct)$ by finding the solutions of \eqref{eqn:Lin_D_SP}. It is therefore  natural to view \eqref{eqn:discreteineq} as an inequality for the maximum value of $\kec$ that would produce growing disturbances for a given value of $\ct=j/\kec t$; disturbances are again restricted to a wedge
\beq
\frac{j}{t}<\ct^*\max_{\ct}\left(\frac{2}{\Real\{\gjk\} }\right)
\eeq where $\ct^*$ is the value of $\ct$ at which the maximum is attained.

Figure \ref{fig:waves_speeds} shows the comparison of the two linearized analyses (continuous and discrete) together with the results of  numerical simulations of the fully nonlinear problem. \cite[To prevent noise triggering the inherent global instability before the disturbance wedge arrives, we used a grid that shifts along with the wave front. More details are given by][]{Lister2010}. We see good agreement between the numerical simulations and the asymptotic results of the linearized model. We also note that in the limit $\kec\ll1$ the results of the discrete and continuous analyses coincide, as should be expected since the characteristic length scales of the problem increase with decreasing $\kec$.

As in the continuous case, we also obtain a prediction, $2\pi\ct/\Imag[g_{j,k}]$, for the number of blocks in the clusters that form at the edge of the wedge, which requires numerical evaluation in general. The results of the discrete and continuous calculations are plotted for a range of spring stiffnesses $\kec$ in figure~\ref{fig:waves_speeds}(\emph{b}) and again can be seen to agree in the limit of large cluster sizes (which is obtained for $\kec\ll1$). The characteristic cluster size resulting from localized disturbances is completely different from the most unstable cluster size predicted by the linear stability analysis of \S\,\ref{sec:linstab}, which is pairwise clusters for all $\kec$. This suggests that localized perturbations may be partly responsible for the relative rarity of pairwise clusters in the simulations shown in figure \ref{fig:nums_ran}. We therefore move on to consider the statistics of cluster size.

\section{Statistics of cluster size}
\label{sec:Stats}

The numerical simulations presented in figure \ref{fig:nums_ran} showed that a range of cluster sizes emerge in a given simulation. This is in contrast to the single preferred cluster size that might be expected from the linear stability analysis, which showed that two-block clusters are always the most unstable. In this section we collate results for a range of values of $\kec$, and examine the statistics of the cluster size.

\begin{figure}
\begin{center}
\mbox{
\begin{tabular}{c}
{\includegraphics[height=0.5\textwidth,angle=0]{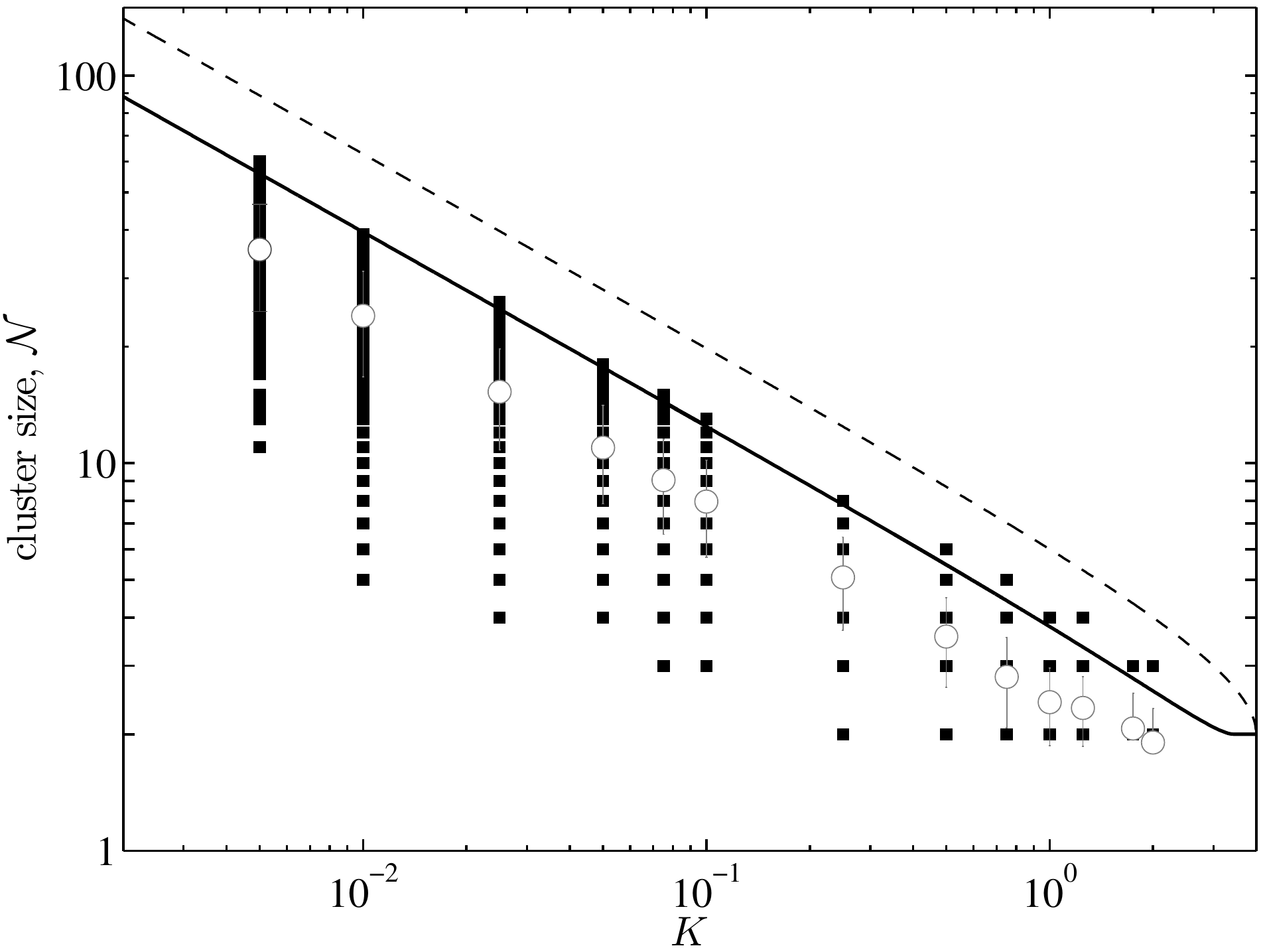}}
\end{tabular}
}
\end{center}
\caption{
The range of cluster sizes observed in numerical simulations with the random initial condition \eqref{eqn:ran_ic} and $\epsilon=10^{-2}$ ($\blacksquare$) for a range of spring stiffnesses $\kec$. The results of one simulation are shown for each value of $\kec$ together with the mean cluster size ($\bigcirc$) and the standard deviation (vertical error bar). Superimposed are the theoretical results for the cluster size at the propagation front caused by a localized disturbance obtained using the  discrete calculation of \S\ref{sec:discretelocalized} (solid curve). The corresponding results for the largest unstable cluster size as predicted by the discrete linear stability analysis of \S\ref{sec:discretelinstab} are also shown (dashed curve). Here $\Ntot=10^4$.
}
\label{fig:clusz_mean_rand}
\end{figure}

The cluster sizes resulting from a single simulation at a number of values of $\kec$ are shown in figure~\ref{fig:clusz_mean_rand}. These show that, for $\kec\ll1$ at least, the mean cluster size appears to scale with $\kec^{-1/2}$. This is a similar scaling to that of the calculated cluster size at the edge of a propagating front, see \eqref{eqn:Nmax:cont}, and also to that of the largest unstable wavelength as calculated by linear stability analysis, \eqref{eqn:discretemaxN}. A more interesting feature of figure \ref{fig:clusz_mean_rand} is that the largest cluster size observed in generic, random initial conditions appears to be similar to that at the propagating front induced by a localized disturbance. This suggests that this cluster size may be `frozen in' as the fronts associated with small localized perturbations pass a given location.

\begin{figure}
\centering
\includegraphics[width=0.7\textwidth,angle=0]{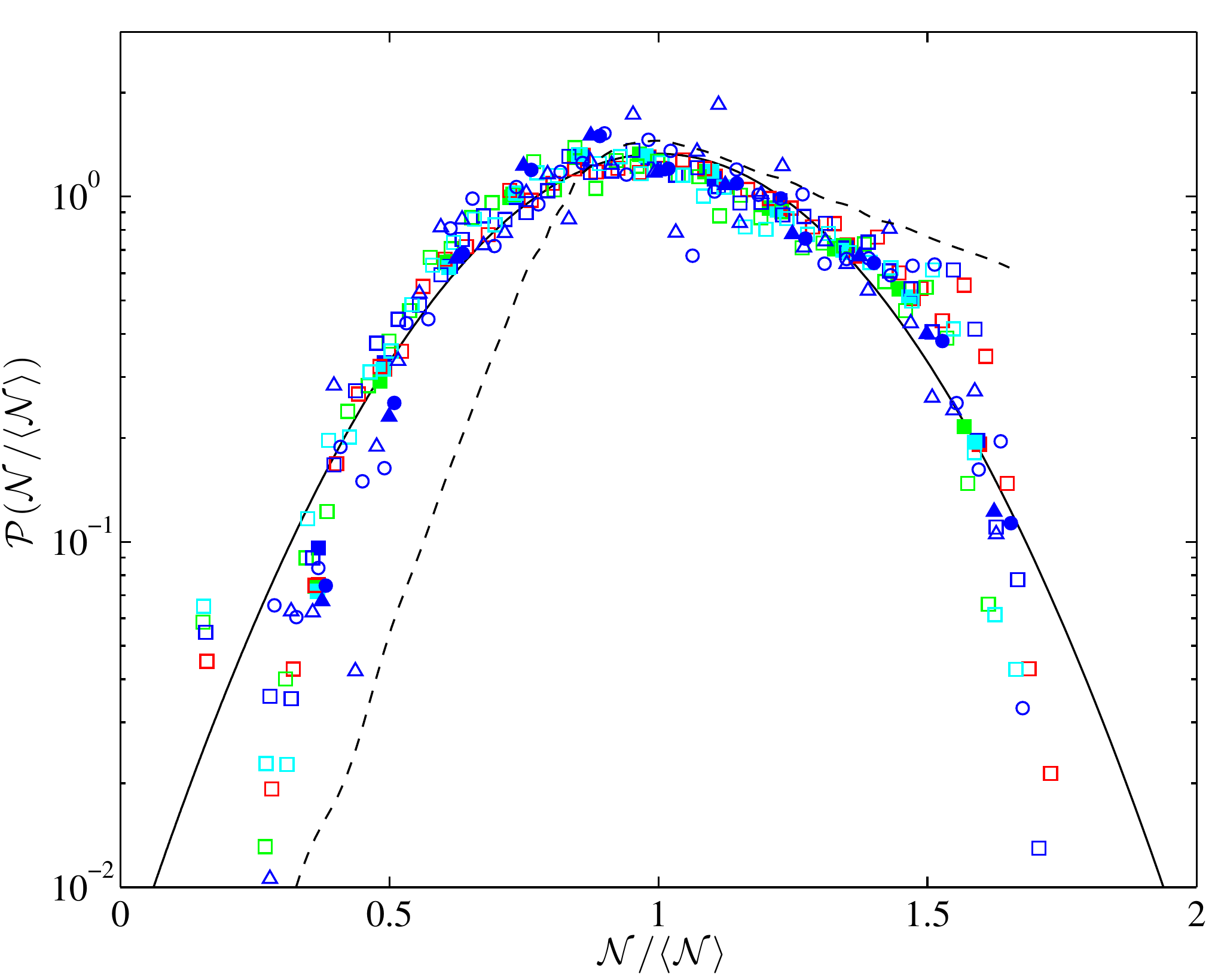}
\caption{Probability distributions of the rescaled cluster size, $\Ns/\Nmean$, computed for a range of sampling distributions, amplitude disturbances and values of $\kec$. Sampling distributions chosen are Uniform($\Box$), Gaussian($\bigcirc$), Gamma ($\bigtriangleup$). Disturbance amplitudes of $\epsilon=10^{-1}, 10^{-2}, 10^{-3}, 10^{-4}$ (cyan, blue, green, red, respectively) were used. Dimensionless spring stiffnesses $K=0.1, \ 0.01$  (closed and open symbols, respectively) were used.
The dashed curve shows the probability distribution found by \cite{Boud2007} from a mean-field aggregation model with a single fitting parameter. (The scaling function $\phi(N/\Nmax)$ presented in Figure 2 of their paper has been rescaled here and plotted as $[\phi/\int_0^1\phi(x)\D x] \Nmean/\Nmax$. The rescaling of the horizontal coordinate, $N/\Nmax\to N/\Nmean$, makes the shape of the distributions comparable; the vertical rescaling ensures that the area under the curve is unity so that the scaling function $\phi$ becomes a probability distribution function.)
} 
\label{fig:stats}
\end{figure}

To test this behaviour more thoroughly, we also considered the variation of the cluster-size distribution with both the size and type of random perturbation to the equilibrium state $h_j=1$. Uniform, Gaussian, and Gamma distributions were used. For each value of $\kec$ and each sampling distribution, 30 simulations were performed from which an averaged distribution could be determined. A probability density function $\Pc$ for cluster sizes $\Ns$ rescaled by the mean cluster size, $\Nmean$, may then be determined from the number of clusters of size $\Ns$, denoted by $n(\Ns)$, according to
\begin{equation} 
\Pc\big({\Ns}/{\Nmean}\big)=n(\Ns)\frac{{\Nmean}^2}{\Ntot}.
\label{eqn:prob}
\end{equation} These averaged probability densities are presented in Figure~\ref{fig:stats}. 

Figure~\ref{fig:stats} shows that, once rescaled in this way, the data collapses well. Furthermore, the observed distribution is very close to a normal distribution with unit mean and standard deviation  $0.3$. (The largest discernible difference is in the tails, which is natural since the normal distribution assigns a non-zero probability to events far away from the mean while our system has well-defined limits on the cluster size.) We note that this distribution appears to be independent of the size of the initial perturbation ($\epsilon$) as well as the distribution from which the initial condition is sampled. In this sense the final cluster-size distribution appears to be universal. Figure~\ref{fig:stats} also compares the cluster-size distribution in our dynamic model of coalescence with the shape of the distribution of sizes found by \cite{Boud2007} in a toy model of sequential aggregation. The two distributions have similar modal values, but the tails of the distribution obtained by \cite{Boud2007} are more skewed than the tails of our nearly symmetric distribution. Given the difference in skewness, the absolute maximum in cluster size they obtain, $\Nmax/\Nmean\approx 1.65$,  is surprisingly close to the effective maximum value, $\Nmax/\Nmean\approx 1.7$, that we obtain.

The appearance of a Gaussian distribution in figure \ref{fig:stats} is, on the face of it, somewhat surprising. The linear stability analysis of \S\,\ref{sec:linstab} suggests that clusters of size $2$ should be prevalent since they are the most unstable clusters for all spring stiffnesses $\kec$. On the other hand, the front propagation caused by localized perturbations leads to clusters of size $\Np(\kec)$ given by \eqref{eqn:Nmax:cont} in the continuous approximation. Combining these two results, we might possibly expect a bimodal distribution; the Gaussian distribution of figure~\ref{fig:clusz_mean_rand} shows that this is not the case. As an alternative hypothesis, \cite{Gat2013} expected that the maximal instability of pairwise clusters should lead to a hierarchy of clusters each containing $2^n$ beams for some integer $n$; we see no evidence of this in our numerical experiments. A third suggestion might be that as clusters of typical size $N$ form they have an effective stiffness $NK$ and are separated by a gap of width $wN$ in which $x\sim x_0/N$, and thus the dimensionless effective spring stiffness $\kec_{\mathrm{eff}}\sim \kec N^4$, so-called `collaborative stiffening' \cite[][]{Bico2004}. If coarsening of clusters progresses until the effective spring stiffness is large enough for the array to be stable (e.g.~$\kec_{\mathrm{eff}}=4$ from the linear stability analysis) then we would expect $N\sim \kec^{-1/4}$. However, we do not see any sign of this in our numerical results. Perhaps the Gaussian distribution represents the result of a sum of many propagating disturbances.

\section{Conclusions \label{sec:Conclusions}}

In this paper we have presented the results of a theoretical investigation of a simplified multi-body elastocapillary system comprising of an array of rigid blocks tethered to their starting positions by linear springs but driven to aggregate by the surface tension of liquid drops placed between the blocks. Studying this simplified system has allowed us to study the dynamics of a system containing up to $\Ntot=10^4$ elements using a model based on the equations of lubrication theory, rather than the kinetic aggregation models that have been used previously \cite[][]{Py2007b,Boud2007}.

The dimensionless parameter that controls the behaviour of this system is the  dimensionless spring stiffness (or elastocapillary number), $\kec$. This spring stiffness governs the stability of an initially uniformly spaced array; for $\kec>4$ the system is stable while for $\kec<4$ the system collapses into a series of clusters whose typical size is also dependent on $\kec$. A linear stability analysis revealed that pairwise clusters are most unstable while a maximum unstable cluster size $\propto\kec^{-1/2}$ exists. Numerical simulations of the uniform array subjected to small random perturbations showed that a range of cluster sizes (other than pairs) are observed and that, while the maximum cluster size does scale with $\kec^{-1/2}$ as expected, the observed pre-factor is significantly smaller than expected. We believe that this maximum cluster size is instead determined by the propagation of fronts through the system that are generated by local perturbations; these fronts have a characteristic cluster size associated with them that agrees well with the maximum cluster size observed numerically. The distribution of cluster sizes appears to be approximately Gaussian suggesting that the interaction of many disturbances acts to smooth out the bimodal distribution that might be expected from the combination of linear stability analysis and front propagation.

From a practical point of view, our results show that sufficiently stiff systems ought to be stable against the formation of clusters, which may be a useful observation in the optimization of the rinsing step in MEMS manufacture. It is also tempting to speculate that some hairy biological systems might be tuned to avoid instability by having sufficiently stiff hairs. For example, it seems feasible that the fine mesh of hairs that is vital for the operation of plastrons, which are used as gills by a number of semi-aquatic insects \cite[see][for example]{Bush2006,Bush2008}, might avoid elastocapillary instability by satisfying the appropriate analogue of the stiffness constraint $\kec>4$ found here.

Although much attention has been focussed on avoiding instability, the formation of clusters can sometimes be a useful tool from the perspective of self assembly \citep[see][for instance]{Pokroy2009}; our results show that noise in the initial conditions can influence the final cluster distribution because this unstable medium is sensitive to the propagation of fronts. Many of these scenarios actually involve a two-dimensional array of elastic elements (rather than the one-dimensional array considered here). In such a geometry we expect the details of the hydrodynamic interactions between elements to be more complicated since the meniscus will, in general, take a more complicated form and the motion of an individual element will be influenced by that of several others. Aside from questions of geometry, an important ingredient present in many of the practical scenarios but absent from our present model is the evaporation of the liquid component. Very rapid evaporation might be expected to stabilize the system against the clustering instability since the liquid should disappear before aggregation has had time to occur. However, it is also possible that the reduction in liquid volume caused by evaporation could reduce the lubrication-induced resistance to clumping that causes very slow collapse of clusters. Which of these effects dominates remains an interesting and important open question.

\section*{Acknowledgements}
KS and DV wish to acknowledge the support of the King Abdullah University of Science and Technology (KAUST; Award No. KUK-C1-013-04), and the John Fell Oxford University Press (OUP) Research Fund. JRL gratefully acknowledges support from Princeton University to cover a sabbatical visit during the Spring semester 2013, and we also acknowledge all those involved in the Oxford--Princeton Collaboration 2013 that prompted some early stages of this work.

\appendix

\section{Asymptotic cluster collapse}
\label{app:asymp_clump}
\begin{figure}
\centering
\mbox{
\begin{tabular}{cc}
\subfigure[]
{\includegraphics[width=0.475\textwidth,angle=0]{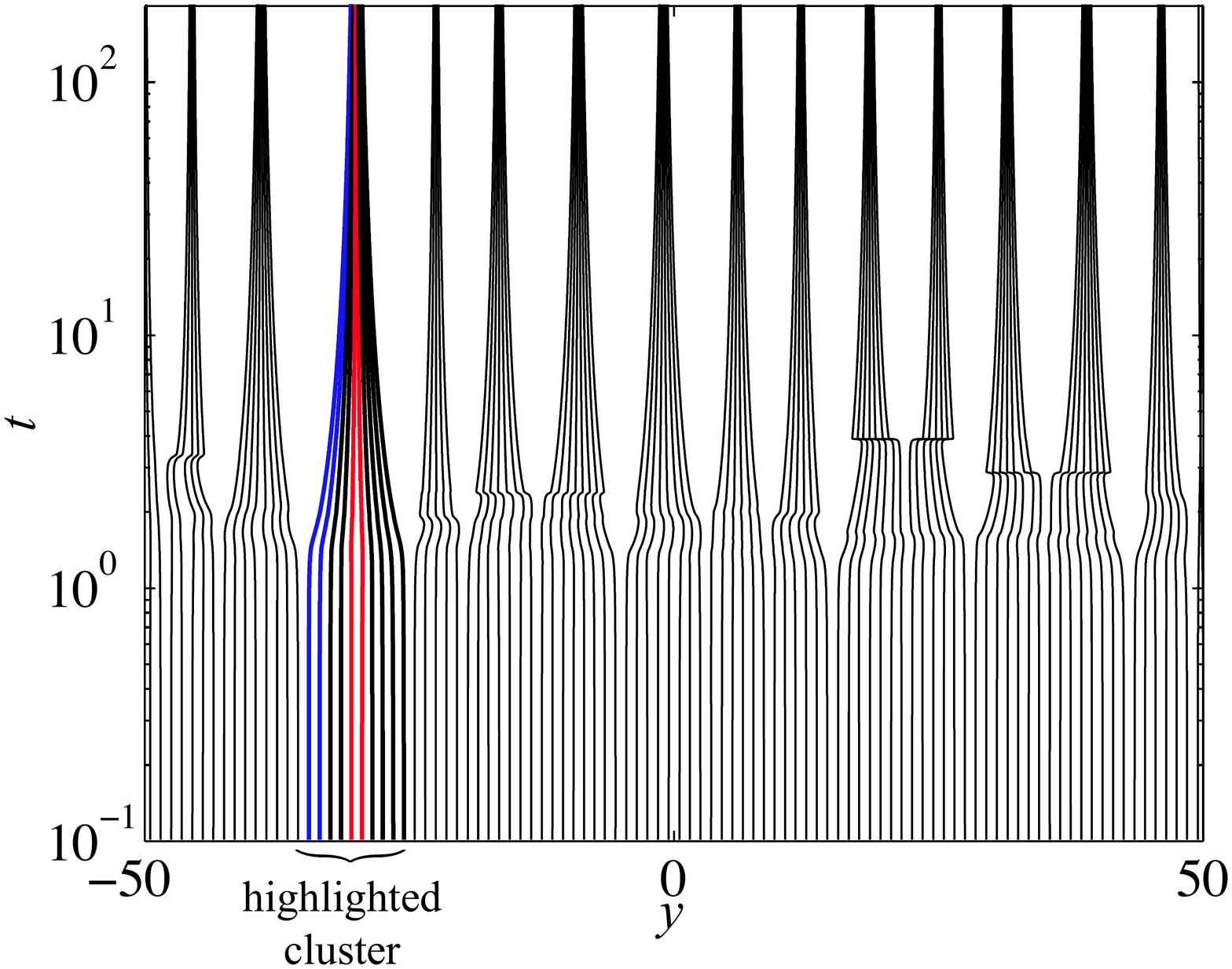}
}
&
\subfigure[]
{\includegraphics[width=0.475\textwidth,angle=0]{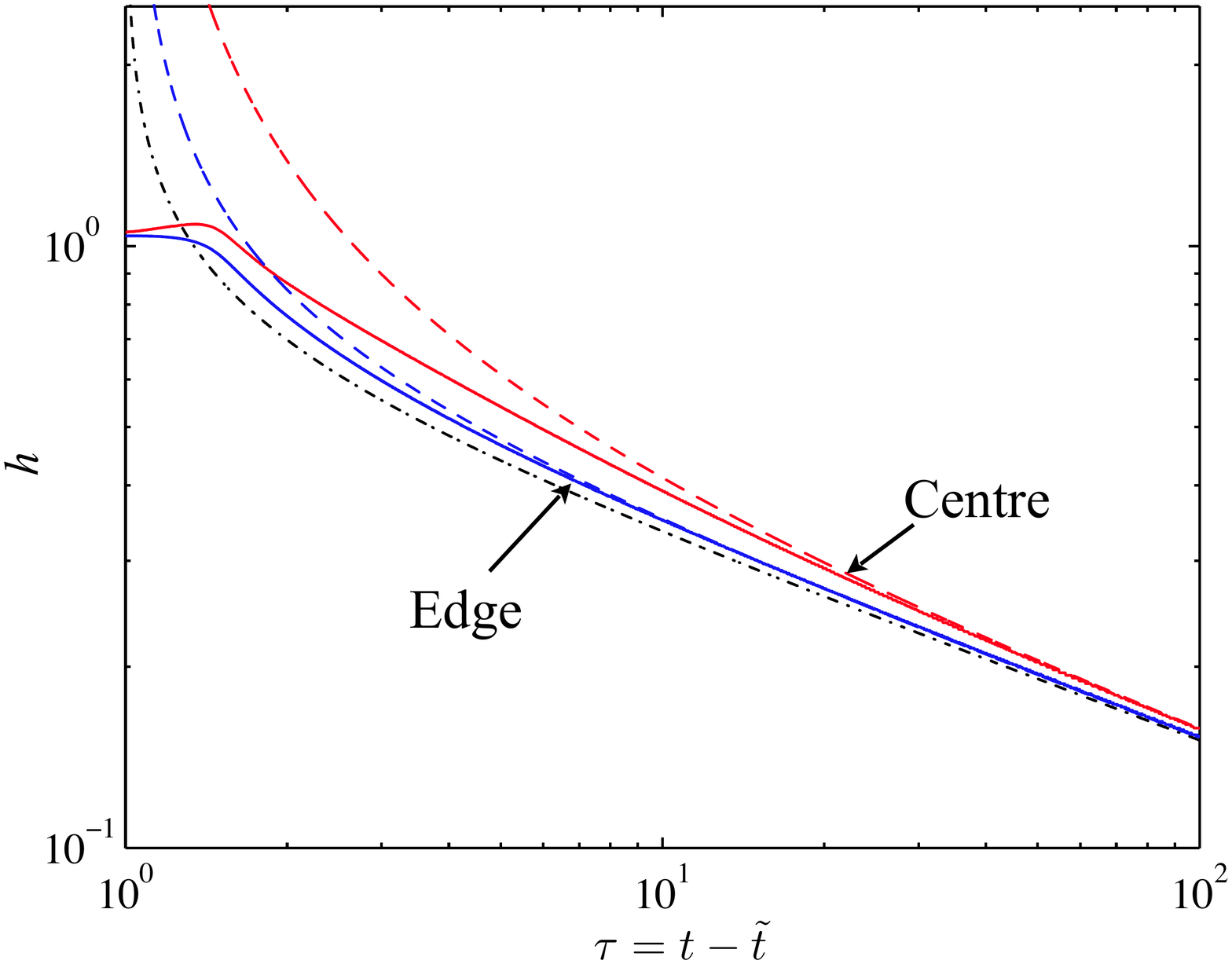}
}
\end{tabular}
}
\caption{The asymptotic evolution of gap widths within a cluster. ({\it a}) The highlighted cluster has $M=10$ blocks. ({\it b}) The solid curves show the evolution of the outermost and central gap widths (blue and red curves, respectively). These are compared with the asymptotic solution from \eqref{eqn:hl} with leading-order $(3\tau)^{-1/3}$ term (dash-dotted curve) and $O(\tau^{-1})$ correction term (dashed curves). (The only fitting parameter is the time offset $\tilde{t}=1.0$; the values of $F_M$ and $F_0$ used in \eqref{eqn:Fl_k} are calculated from the final spacing between clusters.) Here $N=100$ and $\kec=0.1$ while the initial condition was of the form \eqref{eqn:ran_ic} with $\epsilon=0.01$.
}
\label{fig:Clumps}
\end{figure}

In this Appendix, we discuss the dynamics and asymptotic behaviour of the late-time collapse of a single cluster, such as the one highlighted in figure~\ref{fig:Clumps}({\it a}), once the gaps within the cluster are much less than the gaps between adjacent clusters. 

Suppose a cluster of $M$ blocks numbered $1\over 2$ to $M-{1\over2}$ has formed with gaps $h_m\ll1$ for $m=1,\dots,M-1$. The gaps $h_0,h_M$ to the neighbouring clusters are both $O(1)$ and tend towards constant values as each cluster collapses. At leading order, we can thus neglect the viscous resistance between clusters and approximate $F_0$ and $F_M$ by just the capillary attractions $1/h_0^2$ and $1/h_M^2$ and, moreover, can approximate $h_0$ and $h_M$ by the late-time constant asymptotic distance between the cluster centres. (If the cluster is completely isolated, like in the two-block problem of \S\,\ref{subsec:pairs}, then we can simply set $F_0=F_M=0$.)

Let the centre of the cluster be at $y=y_c$. Since the cluster is assumed to collapse and $h_m\to0$ at late times, the position of each block  approaches that of the cluster centre, and $y_m\sim y_c$. Hence, using \eqref{eqn:yi}, the displacement of the $j$th block satisfies  $f_{j+1/2}\sim y_c-(j+\frac{1}{2})$ and, from the force balance \eqref{eqn:EOM1ndim}, $F_{j+1}-F_j\sim 2\kec[y_c-(j+{1\over2})]$. Summing the force balance over the first $m$ blocks, we obtain 
\beq
F_m-F_0=2\kec \sum_{j=0}^{m-1}f_{j+1/2}\sim \kec(2my_c-m^2).
\label{eqn:Fl_F0} 
\eeq
Setting $m=M$ in \eqref{eqn:Fl_F0} yields $y_c\sim M/2+(F_M-F_0)/(2\kec M)$, which shows that the position of the cluster centre is determined by a balance between the net force $F_M-F_0$ on the cluster and the sum of all the restoring spring forces in the cluster. We eliminate $y_c$ from \eqref{eqn:Fl_F0} to obtain 
\beq
F_m\sim \kec m(M-m)+(F_M-F_0)(m/M)+F_0.
\label{eqn:Fl_k} 
\eeq
For example, in the simple case of a completely isolated cluster, $F_0=F_M=0$, we see that $F_m$ has a parabolic variation between zero at the edges and a maximum in the middle.

Lubrication theory shows, via \eqref{eqn:ND_FP}, that the evolution of the gap thickness  $h_m$ depends on the  force $F_m$ according to
\beq
\frac{\dot{h}_m}{h_m^4}=-1+F_m h_m^2.
\label{eqn:hldot}
\eeq
With $F_m$ asymptotically constant and given by \eqref{eqn:Fl_k}, the solution to \eqref{eqn:hldot} satisfies
\beq
h_{m} \sim (3\tau)^{-1/3}+F_m (3\tau)^{-1}
\label{eqn:hl}
\eeq
as $t\to\infty$, where $\tau=t-\tilde{t}$, and $\tilde{t}$ depends on initial conditions. This behaviour is illustrated in figure~\ref{fig:Clumps}({\it b}), by comparing the numerical solution with the asymptotic result \eqref{eqn:hl} for fitted values of $\tilde{t}$ and $F_0$. At leading order all spacings are $(3t)^{-1/3}$, and the next correction is a parabolic variation with $m$ (due to the springs) such that the blocks the middle of the clump haven't collapsed quite as far as those at the edges.

\section{Details of the stationary-phase calculation}

\subsection{ The continuous problem}
\label{app:SP_C}

On letting $c = y/\tau$, the integral in \eqref{eqn:f_st} takes the classic form for application of the method of steepest descents \cite[e.g.][]{Hinch1990,Bender1991}. This method shows that the leading-order asymptotic behaviour in the limit $\tau\gg1$ with $c$ fixed comes from  stationary points $\ell_*$ of the argument $\tau\phi(\ell, c)$ of the exponential term in the integrand. In our case
\begin{equation}
\phi(\ell,c)=\jj \ell c-\ell^{-2},
\label{eqn:phi_kc}
\end{equation}
and the stationary points of $\phi$ are $\ell_*=({2}/{c})^{1/3} \exp\big[(4n{-}3)\jj {\pi}/6\big],\ n=0,1,2$. By deforming the integration contour from the real axis (between $\pm \infty$) to a steepest-descent path, it is found that the relevant saddle points are the pair 
\begin{equation}
\ell_*^{\pm}=\bigg(\frac{2}{c}\bigg)^{1/3}\bigg(\frac{\jj}{2}\pm\frac{\sqrt{3}}{2}\bigg).
\label{eqn:roots_c}
\end{equation}
Evaluating the asymptotic contributions to the integral \eqref{eqn:f_st} from the neighbourhood of this pair of saddle points in \eqref{eqn:roots_c}, and writing $\zh_0(\ell_*^\pm)=I_0\exp(\mp\jj \psi_0)$, we obtain
\begin{align}
\eta(y,\tau)\approx I_0\frac{2^{2/3}}{3^{1/2}\pi^{1/2}}\frac{\tau^{1/6}}{y^{2/3}}\exp\bigg(-\frac{3}{2^{5/3}}y^{2/3}\tau^{1/3}\bigg)
\cos\bigg(\frac{3^{3/2}}{2^{5/3}}y^{2/3}\tau^{1/3}- \psi_0\bigg).
\label{eqn:f_st_sol}
\end{align} The expression for the disturbance field, $H(y,t)$, in \eqref{eqn:statphasecont} can then be retrieved by recalling that $H(y,t)=e^{2t}\eta(y,2Kt)$.

We note that we assumed above that the Fourier transform $\zh_0(\ell)$ exists and is sufficiently smooth near the saddle points; these conditions are satisfied when the initial condition is sufficiently localized, which is certainly the case for \eqref{eqn:LinHIC}. We also note that the above analysis is, in fact, a special case of a general technique used to understand the propagation of fronts into unstable media using only the form of the dispersion relation \citep[see][for a review]{vanS2003}. We present the calculation here both for completeness and because it is a useful starting point for understanding the more complicated analysis required for the discrete problem.

\subsection{The discrete problem}
\label{app:SP_D}

\begin{figure}
\centering
\includegraphics[width=0.95\textwidth,angle=0]{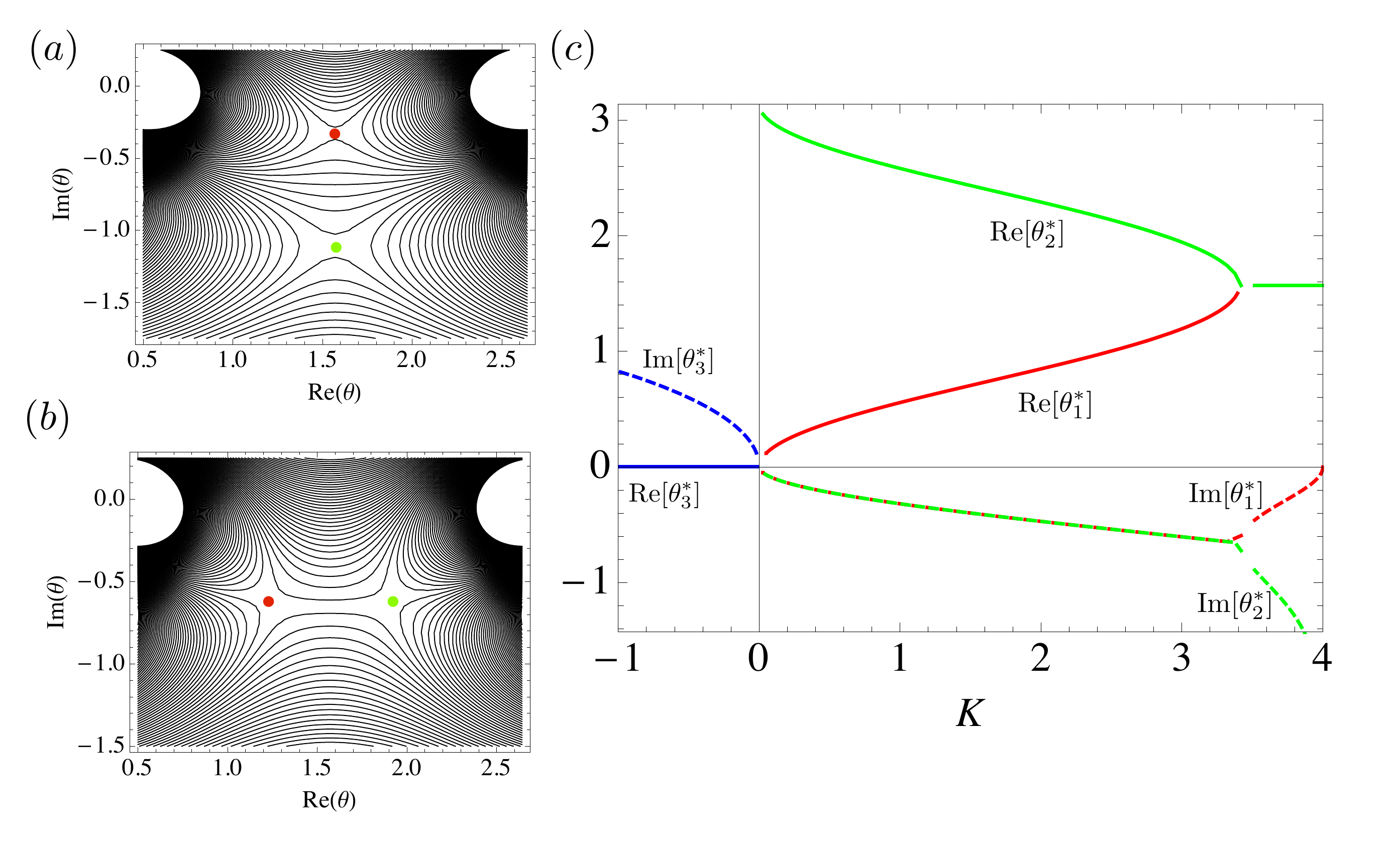}
\caption{ ({\it a},{\it b}) The behaviour of $\Real(g)$ in the complex $\theta$ plane changes for $\ct$ either side of $\ccr = 3^{-3/2}$, because the saddle points move in the complex $\theta$-plane. For ({\it a}) $\ct < \ccr$, the steepest-descent path goes through the higher saddle point (closest to the real axis) whereas in ({\it b}) for $\ct > \ccr$, the path goes through both saddle points. ({\it c}) The dependence of the three roots $\ths_k$ of \eqref{eqn:gpr_app} on the spring stiffness $\kec$. Here real and imaginary parts of the roots correspond to the solid and dashed curves, respectively.} 
\label{fig:waves_roots_thVc}
\end{figure}

The analysis for the discrete problem follows similar steps to those  presented in \S\,\ref{sec:contlocalized} for the continuous problem. We begin seeking  solutions for $H_j(t)$ by making the  transformation
\begin{equation}
H_j=e^{2t}\eta_j,
\label{eqn:Xjzj}
\end{equation}
and defining a rescaled time $\tau=\kec t$ so that \eqref{eqn:LinH} reduces to
\begin{align}
\frac{\D \eta_{j+1}}{\dtau}-{2}\frac{\D \eta_j}{\dtau}+\frac{\D \eta_{j-1}}{\dtau}=2 {\eta_j}.
\label{eqn:Lin_EOM_D}
\end{align}
For simplicity, we assume that $\eta_j=0$ for $\tau<0$ and consider the disturbance produced by a forcing term $\mathcal{F}_j=\delta_{j,0}\delta(\tau)$ in \eqref{eqn:Lin_EOM_D} that is localized in both space and time. Then taking the Laplace transform of \eqref{eqn:Lin_EOM_D}, we obtain
\begin{align}
p \zh_{j+1}-2(p+1) \zh_j+p \zh_{j-1}=\delta_{j,0}.
\label{eqn:Lin_EOM_D_LT}
\end{align}
(A localized initial condition $\eta_j(0)=\delta_{j,0}$ would give $\delta_{j,1}-2\delta_{j,0}+\delta_{j,-1}$ on the right-hand side of \eqref{eqn:Lin_EOM_D_LT}, which complicates the algebra without changing the form of the asymptotic result.)
Seeking a homogeneous solution for $j\ne0$ in \eqref{eqn:Lin_EOM_D_LT} suggests the form $\zh_j\propto m^j$, where $m$ satisfies a certain quadratic equation with roots 
\begin{equation}
m_\pm=1+\xi\pm\big[\xi(\xi+2)\big]^{1/2}, 
\label{eqn:m_xi}
\end{equation} 
where $\xi=1/p$. Note that $m_-=1/m_+$, and $m_+>1$ for $\xi>0$. Requiring the solution of \eqref{eqn:Lin_EOM_D_LT} to be symmetric about $j=0$ and to decay as $j\to\pm\infty$, we deduce that $\zh_{j}=Am^{-|j|}$, with $m=m_+$. The coefficient $A$ is determined from the case $j=0$ in \eqref{eqn:Lin_EOM_D_LT}, with the result that
\begin{align}
\zh_{j}=-\frac{m^{-|j|}}{2p[\xi(\xi+2)]^{1/2}}.
\label{eqn:zhat}
\end{align}
 Taking the inverse Laplace transform of \eqref{eqn:zhat}, we obtain
\begin{align}
\eta_{j}(\tau)=-\frac{1}{4\pi \jj}\int_\gamma \frac{m^{-|j|}e^{p\tau}}{[\xi(\xi+2)]^{1/2}}\frac{\D p}{p},
\label{eqn:zi}
\end{align} where $\gamma$ is the Bromwich inversion contour. The square root in \eqref{eqn:m_xi} is defined to be real and positive for $\xi>0$, and the square root in \eqref{eqn:zi} is defined by analytic continuation with a branch-cut taken along $\xi\in (-2,0)$, equivalent to $p\in(-\infty,-1/2)$.

Note that $p[\xi(\xi+2)]^{1/2}={(1+2p)^{1/2}}$ and $m^{-j}\sim (p/2)^j$ as $p\to0$. Hence the integrand is regular at $p=0$, and we may deform the Bromwich contour to the left in the $p$-plane and collapse it onto the branch cut. We parameterize the two sides of the cut by setting $\cos2\theta=1+\xi$ and $[\xi(\xi+2)]^{1/2}=\jj\sin2\theta$ for $0\leqslant\theta\leqslant\pi$, and find  that $m=\exp(2\jj\theta)$. Using $-\xi=2\sin^2\theta$ and $\upd p/p=-\upd\xi/\xi=-2\,\upd\theta/\tan\theta$, we can then re-express \eqref{eqn:zi} as 

\begin{align}
\eta_j(\tau)=-\frac{1}{4\pi}\int_0^\pi \exp\left(-2\jj j\theta- \frac{\tau}{2\sin^2\theta}\right) \frac{\D \theta}{\sin^{2}\theta} ,
\label{eqn:Lin_D}
\end{align}
which is a convenient form to analyse by the method of steepest descents.

Although $j$ is a discrete variable in the original problem, we can treat it as a continuous variable for the asymptotic analysis. Letting $j/\tau=\ct$,  yields the integral
\begin{align}
\eta_j(\tau)=\frac{1}{4\pi}\int_0^\pi \exp[-g(\theta)\tau] \frac{\D \theta}{\sin^{2}\theta} ,  
\label{eqn:Lin_D2}
\end{align} 
where $g(\theta)=2\jj \ct\theta+1/(2\sin^2\theta)$. In the limit $\tau\to\infty$ with $\ct$ fixed, the solution will be dominated by the neighbourhood of stationary points, $\ths$, corresponding to $g'(\theta)=0$. This yields a cubic equation for $\tan\ths$,
\begin{align}
2\jj \ct\tan^3\ths-\tan^2\ths=1,
\label{eqn:gpr_app}
\end{align} 
with roots $\ths_k$, $k\in(1,2,3)$, and corresponding saddle-point values $g_k=g(\ths_k)$. One root, $\ths_3$ say, is always purely imaginary, and not relevant to the path of integration in \eqref{eqn:Lin_D2}.

We find that $\ccr = 3^{-3/2}$ is a critical value dividing different behaviours of the other two roots, as illustrated in figure~\ref{fig:waves_roots_thVc}({\it a,b}) by contour plots of $\Real[g(\theta)]$ in the complex $\theta$-plane. For $\ct< \ccr$, there are two complex roots $\ths_{1,2}$ of \eqref{eqn:gpr_app}, each with real part $\pi/2$. In this case, the relevant steepest-descent path from 0 to $\pi$ passes through only the saddle point $\ths_1$ closest to the real axis.
For $\ct=\ccr$, the saddle points coincide at $\ths_{1,2} = \pi/2-\jj \coth^{-1}(\sqrt{3})$. For $\ct > \ccr$, the two saddle points have equal imaginary part, and real parts that move symmetrically away from $\pi/2$ as $\ct$ increases. In this case, the steepest-descent path passes through both saddle points, the values of $g(\ths_{1,2})$ are complex conjugates of one another, and both saddles contribute at leading order to the integral.

The leading-order behaviour of \eqref{eqn:Lin_D2} as $\tau\to\infty$ is thus given by
\begin{align}
\eta_j(\tau)\sim \sum_k \frac{\exp[- g(\ths_k)\tau]}{[8\pi(1+2\cos^2\ths_k)\tau]^{1/2}},
\label{eqn:Lin_D_Fnl}
\end{align} 
where we sum over $k=1$ for $\ct< \ccr$ and $k=1,2$ for $\ct> \ccr$, as discussed above. As before, we rescale time and substitute into \eqref{eqn:Xjzj} to obtain equation \eqref{eqn:LinH} for the leading-order behaviour of the disturbance field. 

Using the rescaled wavespeed, $\ct$, as a parameter, we solve for $\ths_k(\ct)$ numerically; the marginal-stability condition \eqref{eqn:discreteineq} then gives an expression $\kec(\ct)=2/\Real[g(\ths_k)]$ for the value of $\kec$ with frontal propagation speed $\ct$. We find that $\ct$ is a decreasing function of $\kec$, as is $c=\ct \kec$ (figure \ref{fig:waves_speeds}{\it a}).  In the limit $\kec\to4$ (the limiting value for instability), we find that $c\to0$ so that no front propagates for systems that are linearly stable, $\kec>4$. The roots $\theta_*$ are plotted as functions of $\kec$ in figure~\ref{fig:waves_roots_thVc}({\it c}). 

Noting that $g(\ths)$ is complex, so that the asymptotic solution \eqref{eqn:Lin_D_Fnl} is oscillatory, we can also predict the cluster size using a similar argument to that in \S\ref{sec:contlocalized} for the continuous case. In general, this argument predicts clusters of size $2\pi \ct/\Imag[g(\ths_k)]$. For the case $\ct< \ccr$, we found that $\Real(\ths_1) = \pi/2$ and thus $\Imag[g(\ths_1)]=\pi\ct$. Hence we predict only pairwise clusters for $4>\kec > \kec(\ccr)\approx 3.408$. For the case $\ct>\ccr$, i.e.~$\kec<\kec(\ccr)$, we calculate $\ths_{1,2}$ numerically, evaluate $\Imag[g(\ths_k)]$ and predict clusters of size larger than two, see figure \ref{fig:waves_speeds}(\emph{b}). The prediction is a continuous function of $\ct$, whereas the actual cluster sizes must be integers. Numerical simulations suggest that if the prediction is not an integer, the cluster size is a mixture of neighbouring integers; in figure \ref{fig:waves_speeds}(\emph{b}) we plot the mean value with vertical errorbars indicating the range of observed cluster sizes.

\bibliographystyle{jfm}
\bibliography{references}
\end{document}